\documentclass[english,journal, doublecolumn]{IEEEtran}
\usepackage[T1]{fontenc}
\usepackage[latin9]{inputenc}
\usepackage{verbatim}
\usepackage{float}
\usepackage{mathtools}
\usepackage{amsmath}
\usepackage{amsthm}
\usepackage{amssymb}
\usepackage{graphicx}

\makeatletter

\floatstyle{ruled}
\newfloat{algorithm}{tbp}{loa}
\providecommand{\algorithmname}{Algorithm}
\floatname{algorithm}{\protect\algorithmname}

\theoremstyle{plain}
\newtheorem{thm}{\protect\theoremname}[section]
\theoremstyle{plain}
\newtheorem{cor}[thm]{\protect\corollaryname}
\theoremstyle{plain}
\newtheorem{lem}[thm]{\protect\lemmaname}

\@ifundefined{date}{}{\date{}}
\usepackage{algorithm}
\usepackage{algpseudocode}
\renewcommand{\fnum@figure}{Fig.~\thefigure}

\@ifundefined{showcaptionsetup}{}{%
 \PassOptionsToPackage{caption=false}{subfig}}
\usepackage{subfig}
\makeatother

\usepackage{babel}
\providecommand{\corollaryname}{Corollary}
\providecommand{\lemmaname}{Lemma}
\providecommand{\theoremname}{Theorem}

\begin{document}
\title{Anytime Minibatch with Delayed Gradients}
\author{Haider Al-Lawati, \IEEEmembership{Student Member, IEEE} and Stark
Draper, \IEEEmembership{Senior Member, IEEE}\thanks{This work was presented in part at the Annual Conference on Information
Sciences and Systems, Baltimore, MD, USA, March 2019 \cite{ciss2019}
 and at IEEE International Conference on Acoustics, Speech, and Signal
Processing, Barcelona, Spain, May 2020 \cite{icassp_ambdg}.\protect \\
This work was supported by Huawei Technologies Canada, the Natural
Science and Engineering Research Council (NSERC) of Canada through
a Discovery Research Grant, and the Omani Government Postgraduate
Scholarship. The authors are with Electrical and Computer Engineering
department, University of Toronto, Toronto, Canada (email: haider.al.lawati@mail.utoronto.ca;
starkdraper@utoronto.ca). Copyright (c) 2020 IEEE. Personal use of
this material is permitted. However, permission to use this material
for any other purposes must be obtained from the IEEE by sending a
request to pubs-permissions@ieee.org.}}
\maketitle
\begin{abstract}
Distributed optimization is widely deployed in practice to solve a
broad range of problems. In a typical asynchronous scheme, workers
calculate gradients with respect to out-of-date optimization parameters
while the master uses stale (i.e., delayed) gradients to update the
parameters. While using stale gradients can slow the convergence,
asynchronous methods speed up the overall optimization with respect
to wall clock time by allowing more frequent updates and reducing
idling times. In this paper, we present a variable per-epoch minibatch
scheme called Anytime Minibatch with Delayed Gradients (AMB-DG). In
AMB-DG, workers compute gradients in epochs of a fixed time while
the master uses stale gradients to update the optimization parameters.
We analyze AMB-DG in terms of its regret bound and convergence rate.
We prove that for convex smooth objective functions, AMB-DG achieves
the optimal regret bound and convergence rate. We compare the performance
of AMB-DG with that of Anytime Minibatch (AMB) which is similar to
AMB-DG but does not use stale gradients. In AMB, workers stay idle
after each gradient transmission to the master until they receive
the updated parameters from the master while in AMB-DG workers never
idle. We also extend AMB-DG to the fully distributed setting. We compare
AMB-DG with AMB when the communication delay is long and observe that
AMB-DG converges faster than AMB in wall clock time. We also compare
the performance of AMB-DG with the state-of-the-art fixed minibatch
approach that uses delayed gradients. We run our experiments on a
real distributed system and observe that AMB-DG converges more than
two times.

\end{abstract}

\section{Introduction\label{sec:Introduction}}

Distributed stochastic optimization has become an important tool in
solving large-scale problems such as those found in modern machine
learning applications. Traditionally, such problems have been solved
using gradient-based methods in a serial manner on a single compute
node. However, training deep neural networks using large datasets
with millions of data points in a reasonable amount of time often
requires parallelizing across multiple machines \cite{45187,NIPS2012_4824,vgg16,Zinkevich:2010:PSG}.
Furthermore, in many online learning systems, such as search engines,
data (e.g., queries) arrive at rates that a single node cannot handle
or process in real time \cite{Dekel:2012:ODO:2188385.2188391}. Although
training deep networks using a single GPU is possible, in many large-scale
training and online learning problems, multiple nodes are deployed
in parallel to cooperatively execute the desired tasks.

In the past decade, a large body of work has proposed and analyzed
various schemes to parallelize stochastic optimization and online
learning problems \cite{45187,Zinkevich:2010:PSG,Dekel:2012:ODO:2188385.2188391,NIPS2011_4247,Bottou04largescale,NIPS2012_4687,5936104,hogwild,amb_iclr,2017arXiv170408227J,Langford:2009:SLF,berkeley_speedup_ml,Lian:2015:APS:nonconvex,nokleby_bajwa,Tsianos2016EfficientDO}.
A common approach is to use a network of nodes arranged according
to a \emph{hub-and-spoke} topology. In this topology, a central node,
known as the \emph{master}, keeps track of and updates the optimization
parameters while the other nodes, known as \emph{workers}, calculate
the gradients of the objective function with respect to individual
data points and submit these gradients to the master.

Distributed optimization methods can be categorized as synchronous
or asynchronous. In synchronous methods, the master waits for all
workers to submit their results before updating the optimization parameters.
As a result, the speed of the computation in synchronous methods is
limited by the speed of the slowest workers known as ``stragglers''
\cite{TAS,stragglers_root_cause,Zaharia:2008:IMP}. To alleviate the
straggler effect, asynchronous methods have been proposed in which
the master does not wait for all workers. Instead, the master updates
the optimization parameters using ``stale'' (i.e., delayed) gradients.

Recently, we introduced a novel synchronous technique called \emph{Anytime
Minibatch} (AMB) in \cite{amb_iclr}. Our main idea was to fix the
per-epoch compute time across all workers rather than fixing the job
size as is typical. Since workers progress at different rates, fixing
the compute time results in a variable amount of work being completed
by each worker in each epoch. A drawback of AMB is that when workers
submit their results to the master, they idle while waiting to receive
back the updated optimization parameters. When this wait time is long,
the wall time convergence can be greatly prolonged. To alleviate this
deficiency, in this paper we propose \emph{Anytime Minibatch with
Delayed Gradients} (AMB-DG). In AMB-DG, workers keep processing data
points and calculate gradients at all times while the master updates
the parameters using delayed gradients. We analyze the performance
of AMB-DG for a convex smooth objective function under the assumption
of fixed gradient staleness. Under these assumptions, we prove that
the expected regret bound attained is $O(\sqrt{{m}})$ where $m$
is the number of samples observed across all nodes. This bound is
well-known to be the optimal regret bound achievable by gradient-based
methods on arbitrary convex objective functions \cite{Dekel:2012:ODO:2188385.2188391}.
We also show that under the same assumptions the expected convergence
rate (i.e., optimality gap) achieved is $O(1/\sqrt{m})$. Our theoretical
derivations show that asymptotically in the number of data points
observed, the impact of the fixed delay on convergence is negligible
for smooth convex loss functions. We compare the performance of AMB-DG
with that of AMB and observe that under long communication delays
AMB-DG converges faster than AMB in terms of wall clock time. We also
implement AMB-DG on the SciNet high performance computing platform
and compare the performance of AMB-DG with existing asynchronous delayed-gradient-based
methods. AMB-DG converges almost $2$ times faster in our examples.

The remainder of the paper is organized as follows. In Sec. \ref{sec:RelWork},
we review the literature. In Sec. \ref{sec:AMBDG-Scheme}, we present
the system model and the algorithmic description. The convergence
analysis and theorems are presented in Sec. \ref{sec:ConvRes}. Detailed
proofs are deferred to Appendix \ref{appendix:Proofs}. In Sec. \ref{sec:NumRes},
we present numerical results illustrating the performance of AMB-DG.
Finally, Sec. \ref{sec:Conc} summarizes and concludes the paper.

\section{Related Work\label{sec:RelWork}}

Asynchronous distributed optimization has been studied at least since
the seminal work of Tsitsiklis et al. \cite{Tsitsiklis_1986}. In
recent years, asynchronous schemes have played a key role in solving
large-scale machine learning and online optimization problems. For
instance, Google's \emph{distbelif} \cite{NIPS2012_4687} was one
of the earliest implementation of asynchronous distributed optimization
for large-scale training in machine learning. In general, the literature
pertaining to asynchronous distributed optimization and online learning
focuses on two aspects: algorithmic development and convergence analysis.

Developing schemes for asynchronous distributed optimization has been
an active area of research in recent years. Examples of such algorithms
include \cite{Zinkevich:2010:PSG,hogwild,Langford:2009:SLF,NIPS2013_4939}.
In \cite{NIPS2013_4939}, the authors study parallel optimization
with sparse data. They propose an asynchronous dual averaging approach
termed AsyncDA. In AsyncDA, the master keeps and updates the dual
variable while each worker keeps its own primal optimization variable.
Asynchronous methods use stale gradients, as was mentioned earlier.
While asynchronous schemes can tolerate gradient staleness, large
staleness can slow convergence. This is known as the \emph{stale gradient
problem}. For instance, in \cite{Langford:2009:SLF}, gradient staleness
scales linearly in the number of workers. Thus using larger network
may have an undesirable impact on convergence. To limit the negative
impact of staleness, the authors in \cite{NIPS2013_4894} propose
a stale-gradient-based scheme and control the staleness in a spoke-and-hub
setting by forcing fast workers to wait for slow ones. One of the
challenges encountered in distributed optimization is dealing with
the communication load. When the number of workers increases, worker-master
communication may become a bottleneck. To reduce the communication
load and alleviate the bottleneck, techniques that compress the information
before transmission have been proposed and analyzed such as gradient
quantization \cite{alistarh2017qsgd,NIPS2018_7752,pmlr-v80-bernstein18a}
and gradient sparsification (i.e., randomly dropping out coordinates
of the gradient and amplifying the rest) \cite{NIPS2018_7405}.

In another thread of works, convergence analyses of existing asynchronous
methods have been developed. In \cite{NIPS2011_4247}, the authors
analyze the convergence of distributed optimization of smooth convex
loss functions using delayed gradients under the assumption of bounded
gradient delay. They prove that the effect of gradient delay is negligible
asymptotically in the number of processed data points and that the
optimal rate is achievable. In \cite{Feyzmahdavian2015AnAM}, the
convergence analysis is extended to general regularized convex cost
functions. In \cite{Lian:2015:APS:nonconvex}, the authors study the
convergence of stochastic gradient methods for nonconvex objective
functions and establish an ergodic convergence rate of ${\cal O}(1/\sqrt{T})$,
where $T$ is the number of iterations.

\section{AMB-DG Scheme\label{sec:AMBDG-Scheme}}

In this section, we introduce AMB-DG. We first introduce the system
model and then describe the algorithm. The pseudocode of the scheme
is found in Appendix \ref{app:AMB-DG-Pseudocode}.

\subsection{System Model\label{sec:System-Model}}

In this paper we study a synchronized network consisting of a master
and $n$ workers. Processing proceeds in epochs, each of duration
$T_{p}$ seconds. Epochs are indexed by $t\in\mathbb{Z}_{+}$. The
$t$-th epoch therefore ends at time $tT_{p}$. At the end of epoch
$t$ workers synchronously and in parallel transmit messages to the
master. Worker $i$ transmits message $m_{i}(t)$. All worker-to-master
transmissions take $\tfrac{1}{2}T_{c}$ second. Based on its current
state and the $n$ messages received, the master updates its estimate
of the optimization parameters. The $t$-th update of the optimization
parameter is denoted by $w(t+1)$. We assume the master\textquoteright s
update occurs instantaneously. Therefore the master\textquoteright s
$t$-th update occurs at time $tT_{p}+\tfrac{1}{2}T_{c}$. The master
then broadcasts its $t$-th parameter update to all workers. This
is done in parallel and all messages take $\tfrac{1}{2}T_{c}$ second
to reach each worker. Therefore, each worker receives $w(t+1)$ at
time $tT_{p}+T_{c}$. Naturally, depending on the relative durations
of $T_{p}$ and $T_{c}$ the workers will be basing their calculation
on more or less stale information. The ratio $\tau=\lceil T_{c}/T_{p}\rceil\in\mathbb{Z_{+}}$
plays an important role in the ensuing derivation. We term this ratio
the ``staleness parameter''. Note that $T_{p}$ and $T_{c}$ are
both deterministic; hence $\tau$ is deterministic, too. The staleness
parameter, $\tau$, represents the number of updates the master makes
to the optimization parameter from the time when it calculates and
broadcasts $w(t)$ until it receives gradients with respect to the
same $w(t)$.

\begin{figure}
\begin{centering}
\includegraphics[scale=0.35]{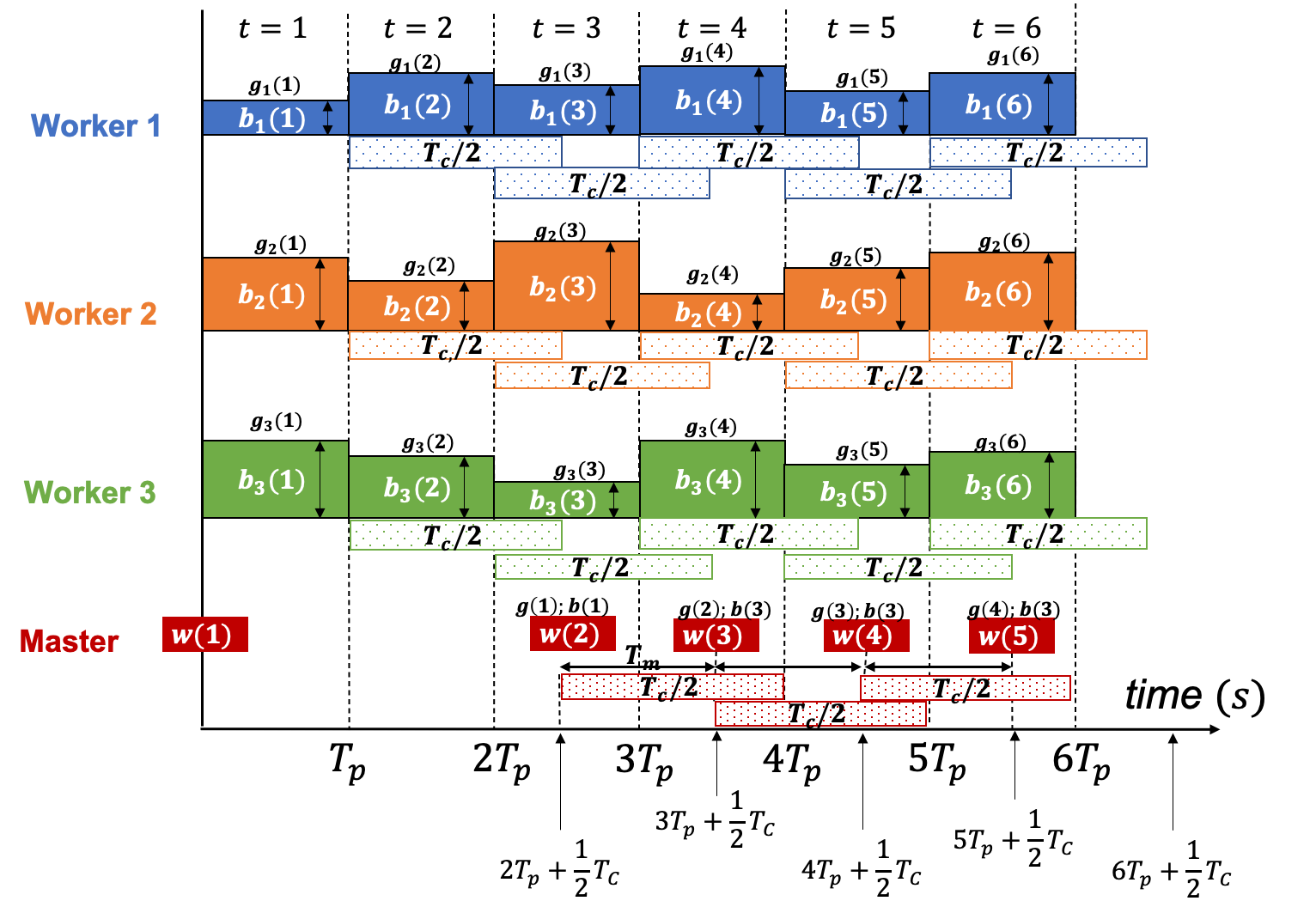}
\par\end{centering}
\centering{}\caption{\label{fig:AMBDG-system-model}AMB-DG system model}
\end{figure}

The above description is diagrammed in Fig. \ref{fig:AMBDG-system-model}.
At the core of our interest in this system is the fact that workers
may be operating at different computing speeds and therefore may process
different amounts of data in each processing epoch. In epoch $t$
worker $i$ processes $b_{i}(t)$ data points. These data are indexed
as $x_{i}(t,s)$ where $s\in[b_{i}(t)]$. The total data processed
in epoch $t$ is denoted $b(t)=\sum_{i=1}^{n}b_{i}(t)$. In Fig. \ref{fig:AMBDG-system-model}
the variable-sized mini-batches are indicated by the different heights
of the bar plots associated with each worker. Fig. \ref{fig:AMBDG-system-model}
illustrates a system with three workers.

\subsection{Algorithmic Description\label{subsec:Algo-desc}}

Our objective is to solve the following optimization problem

\begin{equation}
\underset{w\in\mathcal{{W}}}{\text{{minimize\,}}}F(w)\text{ {\it {\rm where}} }F(w):=\mathbb{E}_{P}\big[f(w,x)\big].\label{eq:optProb-1}
\end{equation}

\noindent In (\ref{eq:optProb-1}) $f:\mathcal{W}\times\mathcal{X}\rightarrow\mathbb{R}$
and the expectation is taken with respect to some (unknown) distribution
$P$ over the set $\mathcal{X}\subseteq\mathbb{R}^{d}$. In particular,
we are interested in solving for $w^{*}={\rm argmin}_{w\in{\cal W}}F(w)$.

The algorithm we study in the context of the above system model is
dual averaging \cite{Nesterov:2009:dual_average,Xiao:NIPS2009_dual_avg}.
We first describe the distributed dual averaging without stale gradients
(i.e., when $T_{c}=0$). Then, we explain the delayed-gradient-based
variant.

Dual averaging is an iterative gradient-based algorithm. In each epoch
$t$, the system updates two parameters: the primal variable (a.k.a.
the optimization parameter) $w(t)$ and the dual variable $z(t)$.
In the distributed implementation of this algorithm, the master tracks
and updates these parameters while workers calculate the gradients.
The system initializes $z(1)=0\in\mathbb{R}^{d}$ and $w(1)={\rm argmin}_{w\in{\cal W}}\psi(w)$
where $\psi(w)$ is a proximal function.

Next, we introduce the gradient computation and parameter update phases
of AMB-DG algorithm. First in the absence of gradient staleness, i.e.,
when $T_{c}=0$, then we explain how to extend the algorithm to cater
for stale gradients.

\subsubsection*{Gradients computation}

In each epoch $t$, worker $i$ calculates $b_{i}(t)$ gradients of
$f(w,x)$ with respect to the optimization parameters $w(t)$ and
data points $\{x_{i}(t,s)\},s\in[b_{i}(t)]$. The sum of gradients
calculated by worker $i$ at the end of epoch $t$ is

\begin{equation}
g_{i}(t)=\sum_{s=1}^{b_{i}(t)}\nabla_{w}f\big(w(t),x_{i}(t,s)\big),\label{eq:gradCompNoStale}
\end{equation}

\noindent where the $x_{i}(t,s)$ are sampled by worker $i$ in epoch
$t$ in an independent and identically distributed (i.i.d.) manner
from $P$. At the end of each gradient computation epoch, each worker
$i$ sends message $\mbox{\ensuremath{m_{i}(t):=\big(g_{i}(t),b_{i}(t)}\ensuremath{\big)}}$
to the master and proceeds immediately to the next epoch.

\subsubsection*{Parameters update}

The master calculates the average of the received gradients $g(t)=\frac{1}{b(t)}\sum_{i=1}^{n}g_{i}(t)$,
and updates $z(t)$ and $w(t)$ using the following dual averaging
update rules

\begin{equation}
z(t+1)=z(t)+g(t),\label{eq:dual_var-1}
\end{equation}

\begin{equation}
w(t+1)=\arg\min_{w\in\mathcal{W}}\Big\{\big\langle z(t+1),w\big\rangle+\frac{1}{\alpha(t+1)}\psi(w)\Big\},\label{eq:primal_var-1}
\end{equation}

\noindent where $\alpha(t)$ is a nonincreasing sequence of \emph{step
sizes}. Since we work in the Euclidean space, a typical choice of
$\psi(w)$ is $\lVert w\rVert^{2}$. The master broadcasts $w(t+1)$
to all workers immediately. The workers use the updated optimization
parameter to calculate the next set of gradients.

\subsubsection*{Stale gradients}

Next, we consider the case when $T_{c}\ne0$. To simplify our analysis,
we assume that $T_{c}$ is an integer multiple of $T_{p}$ so that
$\tau=T_{c}/T_{p}.$ This is possible by choosing the appropriate
$T_{p}$. Note that for short communication delay or when $T_{c}$
is smaller than $T_{p}$, $\tau\le1$. In this case, we can use AMB
instead of AMB-DG as workers' idle time between two consecutive epochs
is relatively small and the benefits of using non-stale gradients
can result in faster convergence despite the short intermittent idle
times. In our analysis, we are interested in operating in the large
communication delay regimes. Large communication delay has been observed
to be a major bottleneck in large scale distributed computing. In
this situation, slow epoch-wise convergence due to delayed gradients
is compensated by a faster rate of parameters update. Furthermore,
since $\lceil T_{c}/T_{p}\rceil\ge T_{c}/T_{p}$ with equality if
and only if $T_{c}$ is an integer multiple of $T_{p}$, setting $T_{p}$
such $T_{c}$ is an integer multiple of $T_{p}$ ensures that workers
will immediately use the new information; i.e., a new compute epoch
starts as soon as updated parameters arrive. Also, since we are interested
in a large communication delay regime, such a choice of $T_{p}$ is
not very limiting.

We assume the use of a non-blocking communication protocol. That is,
after each gradient computation epoch ends, workers send their messages
to the master and immediately start a new round of computation while
communication proceeds in the background. Note that workers receive
the first update from the master at time $T_{p}+T_{c}=(\tau+1)T_{p}$.
Hence, worker $i$ uses $w(1)$ to calculate gradients $g_{i}(t)$
for epochs $1\le t\le\tau+1$. In other words,

\[
g_{i}(t)=\sum_{s=1}^{b_{i}(t)}\nabla_{w}f\big(w(1),x_{i}(t,s)\big).
\]

\noindent On the other hand, since workers receive $w(t+1)$ at time
$tT_{p}+T_{c}=(t+\tau+1)T_{p}$, then for epochs $t\ge\tau+2$, each
worker $i$ calculates $g_{i}(t)$ as

\[
g_{i}(t)=\sum_{s=1}^{b_{i}(t)}\nabla_{w}f\big(w(t-\tau),x_{i}(t,s)\big).
\]

\noindent Therefore, the master calculates $g(t)$ as

\begin{equation}
g(t)=\sum_{i=1}^{n}\sum_{s=1}^{b_{i}(t)}\nabla_{w}f(w(t-\tau),x_{i}(t,s)),\label{eq:delayedGrad}
\end{equation}

\noindent where for $t-\tau\leq0$, $w(t-\tau):=w(1)$, the initial
value.

Figure \ref{fig:AMBDG-system-model} illustrates the delayed gradient
model when the communication time $T_{c}=3T_{p}$. Workers calculate
gradients with respect to $w(1)$ in the first four epochs and send
their messages to the master at times $tT_{p}$ for $t=1,2,3,4$.
The master uses $z(2)=g(1)=\sum_{i=1}^{n}g_{i}(1)$ to calculate $w(2)$
and broadcasts it to all workers while it uses delayed gradients $g(2)$,
which are calculated with respect to $w(1)$, to update $z(3)$ before
calculating $w(3)$. Worker $i$ receives $w(2)$ at time $4T_{p}$
and uses it to calculate $g_{i}(5)$. The master uses $g_{i}(5)$
in turn to calculate $z(6)$ and hence $w(6)$ at time $5T_{p}+\tfrac{1}{2}T_{c}$.
Since $T_{c}=3T_{p}$, the staleness parameter in this example is
$\tau=3$. Observe that the master calculates $w(6)$ using gradients
calculated with respect to $w(2)$ instead of $w(5)$, so gradient
staleness $5-2=3=\tau$. Except for the first $\tau+1=4$ epochs,
the master calculates $w(t)$ for $t\ge\tau+2=5$ using gradients
calculated with respect to $w(t-\tau)=w(t-3)$.

We note that although we use dual averaging as an algorithmic workhorse,
AMB-DG can be implemented using other gradient-based algorithms as
well. The motivation behind using dual averaging stems from the following.
Dual averaging has been shown to asymptotically achieve optimal regret
bound when applied in distributed optimization using delayed gradients
and fixed minibatch size \cite{NIPS2011_4247}. Furthermore, in \cite{amb_iclr},
we proved that when using dual averaging in a synchronous distributed
setting with variable minibatch per epoch with no gradient staleness,
we can achieve optimal regret bound asymptotically. Hence, we anticipate
that using dual averaging with delayed gradients and variable minibatch
per epoch could yield the desired convergence rate.

\section{Convergence Results\label{sec:ConvRes}}

In this section, we develop convergence results for AMB-DG in terms
of expected regret and expected convergence rate. The latter is also
known as the optimality gap. Regret is the standard metric of performance
measurement in the online learning literature while optimality gap
is the standard metric in stochastic optimization. Let $R(T)$ and
$G(T)$, respectively, denote the regret and the optimality gap after
$T$ epochs (i.e., parameters updates). Then,

\begin{equation}
R(T)=\sum_{t=1}^{T}\Big[f\big(w(t+1),x(t+1)\big)-f(w^{*},x(t+1))\Big],\label{eq:regret}
\end{equation}

\noindent while

\begin{equation}
G(T)=F(\hat{w}(T))-F(w^{*}),\label{eq:optGap}
\end{equation}

\noindent where $w^{*}={\rm argmin}_{w\in{\cal W}}F(w)$ and $\hat{w}(T)=\frac{1}{T}\sum_{t=1}^{T}w(t+1)$
is the time-average of the optimization variable over the $T$ epochs.

Before turning to our performance analysis, we first state the main
assumptions used in our derivations.

\subsection{Preliminaries\label{subsec:Preliminaries}}

As was stated in Sec. \ref{sec:AMBDG-Scheme}, we are interested in
solving the online stochastic optimization problem of (\ref{eq:optProb-1}).
We assume that the feasible set $\mathcal{W}$ is closed and convex
and that an optimal solution, $w^{*}\in{\cal W}$, exists. We assume
that $f(w,x)$ is differentiable and convex in $w$ for every $x$.
Hence, $F(w)$ is also convex and differentiable and the gradient
of the objective function is $\nabla F(w):=\mathbb{E}_{P}\big[\nabla f(w,x)\big]$.
We assume that $F(w)$ and $\nabla f(w,x)$ are both Lipschitz continuous
with parameters $J$ and $L$, respectively. In other words,

\begin{equation}
\lvert F(w_{1})-F(w_{2})\rvert\le J\,\,\lVert w_{1}-w_{2}\rVert\quad\forall w_{1},w_{2}\in\mathcal{W},\label{eq:G-Lipschitz}
\end{equation}

\noindent and for all $w_{1},w_{2}\in{\cal W}$ and for all $x\in{\cal X}$

\begin{equation}
\lVert\nabla f(w_{1},x)-\nabla f(w_{2},x)\rVert\le L\lVert w_{1}-w_{2}\rVert\label{eq:L-Lipschitz}
\end{equation}

\noindent where $\lVert.\rVert$ is the $l_{2}$ norm. The smoothness
property in (\ref{eq:L-Lipschitz}) implies that

\begin{equation}
f(w_{2},x)\le f(w_{1},x)+\langle\nabla f(w_{1},x),w_{2}-w_{1}\rangle+\frac{L}{2}\lVert w_{1}-w_{2}\rVert^{2}.\label{eq:smoothProp}
\end{equation}

We also assume the following variance bound, for all $w\in\mathcal{W},\mathbb{E}_{x}\big[\big\lVert\nabla f(w,x)-\nabla F(w)\big\rVert^{2}\big]\le\sigma^{2}$,
for some $\sigma\ge0$. Furthermore, the proximal function, $\psi(w)$,
in the primal variable update rule (\ref{eq:primal_var-1}) is $1$-strongly
convex. This means that $\psi$ satisfies

\begin{equation}
\psi(w_{2})\ge\psi(w_{1})+\langle g,w_{2}-w_{1}\rangle+\frac{1}{2}\lVert w_{2}-w_{1}\rVert^{2}
\end{equation}

\noindent for all $w_{1},w_{2}\in\mathcal{W}$ and for all $g\in\partial\psi(w)$
where $\partial\psi(w)$ is the set of sub-gradients defined as $\partial\psi(w):=\left\{ g\in\mathbb{R}^{d}\vert\psi(w_{2})\ge\psi(w_{1})+\langle g,w_{2}-w_{1}\rangle,\forall\,w_{2}\in\text{dom}(\psi)\right\} $
.

Furthermore, for the optimal solution, $w^{*}$, we assume that $\psi(w^{*})\le C^{2}/2$
and that for all $w\in\mathcal{W},$ $D_{\psi}(w^{*},w)\le C^{2}$,
for some $C\in\mathbb{R}$. In these definitions, $D_{\psi}(w^{*},w)$
is the Bregman divergence between $w^{*}$ and $w$ defined as

\begin{equation}
D_{\psi}(w^{*},w):=\psi(w^{*})-\psi(w)-\big\langle\nabla\psi(w),w^{*}-w\big\rangle.\label{eq:bregman}
\end{equation}

\subsection{\label{subsec:Regret-and-Convergence}Regret and Convergence Analysis}

We analyze the performance of AMB-DG in terms of the regret achieved
after $T$ epochs. The $b_{i}(t)$ are modeled as i.i.d. random variables.
Let $\bar{b}$ be the expected minibatch per epoch. That is, $\mathbb{E}[b(t)]=\bar{b}$.
Then $\mathbb{E}[b(t)]=\mathbb{E}\big[\sum_{i=1}^{n}b_{i}(t)\big]=n\bar{b}$.
Therefore, $\mathbb{E}[b_{i}(t)]=\bar{b}/n$ for all $i\in[n]$ and
for all $t\in[T]$. Furthermore, assume that $b(t)\ge\hat{b}$. Let
$m$ be the expected total number of samples observed from epoch $1$
to $T$, i.e., 

\begin{equation}
m=\mathbb{E}\Big[\sum_{t=1}^{T}b(t)\Big]=\sum_{t=1}^{T}\sum_{i=1}^{n}\mathbb{E}[b_{i}(t)]=T\bar{b}.\label{eq:sample_path}
\end{equation}

We assume that all gradients received are delayed by $\tau$ steps
where $\tau$ is a nonnegative integer.

\begin{align}
\mathbb{E}\big[R(T)\big]\coloneqq\mathbb{E}\bigg[\sum_{t=1}^{T}\sum_{i=1}^{n}\sum_{s=1}^{b_{i}(t)}\Big[f & (w(t+1),x_{i}(t+1,s))\nonumber \\
 & -f(w^{*},x_{i}(t+1,s))\Big]\bigg].\label{eq:cond_regret}
\end{align}

We next state the main theoretical results of our work. The theorem
bounds the expected regret and the corollary bounds the optimality
gap.

\begin{thm}
\label{thm:reg_bnd}Assume that the master receives a set of gradients
over $T$ epochs during which the expected number of data points workers
sample from $P$ in an i.i.d. manner is $m$. Let $\hat{b}\ne0$ and
$\bar{b}$ be as defined above. Choosing $\alpha(t)^{-1}=L+\sqrt{(t+\tau)/\bar{b}}$,
the expected regret for AMB-DG with deterministic common gradient
delay $\tau$ across all master-worker links in each epoch is
\end{thm}
\noindent 
\begin{align}
\mathbb{E}\big[R(T)\big]\le\bar{b}\frac{C^{2}}{2}\Big(L+\sqrt{(T+1+\tau)/\bar{b}}\Big) & +2\tau JC\bar{b}\nonumber \\
+\,2LJ^{2}(\tau+1)^{2}\bar{b}^{2}\big(1+\log T & \big)+\frac{\bar{b}}{\hat{b}}\sigma^{2}\sqrt{m}.\label{eq:regret_bound}
\end{align}
In both Theorem \ref{thm:reg_bnd} and Corollary \ref{cor:opt_gap},
next, $J$ and $L$ are, respectively, the Lipschitz constants of
$F(w)$ and $\nabla f(w,x)$, $\sigma^{2}$ is the bound on the variance
of $\nabla f(w,x)$ and $C$ is a constant that satisfies that for
all $w\in\mathcal{W},$ the Bregman divergence $D_{\psi}(w^{*},w)\le C^{2}$.

Let $\hat{w}(T)$ be the average of the optimization variable over
$T$ epochs, i.e., 
\begin{equation}
\hat{w}(T)=\frac{1}{T}\sum_{t=1}^{T}w(t+1).\label{eq:wtd_output_avg}
\end{equation}
Using $\hat{w}(T)$ the expected optimality gap is
\begin{equation}
\mathbb{E}\Big[G(T)\Big]\coloneqq\mathbb{E}\bigg[F\Big(\frac{1}{T}\sum_{t=1}^{T}w(t+1)\Big)-F(w^{*})\bigg].\label{eq:exp_opt_gap}
\end{equation}

Using the results from Theorem \ref{thm:reg_bnd} and the definition
of the conditional expectation of the optimality gap in (\ref{eq:exp_opt_gap}),
we develop the following corollary on the upper bound of the expected
optimality gap achieved by AMB-DG.
\begin{cor}
\noindent \label{cor:opt_gap} The expected optimality gap for the
AMB-DG scheme is
\begin{align}
\mathbb{E}\Big[G(T)\Big]\le & \bar{b}\bigg(\frac{C^{2}}{2m}\Big(L+\sqrt{(T+1+\tau)/\bar{b}}\Big)+\frac{2\tau JC}{m}\nonumber \\
 & +2LJ^{2}\frac{(\tau+1)^{2}\bar{b}\big(1+\log T\big)}{m}+\frac{\sigma^{2}}{\hat{b}\sqrt{m}}\bigg).\label{eq:opt_gap}
\end{align}
\end{cor}
The proofs for Theorem \ref{thm:reg_bnd} and Corollary \ref{cor:opt_gap}
are provided in Appendix \ref{appendix:Proofs}. The proof techniques
exploits the convexity of $F$ and the Lipschitz continuity of $\nabla F$
to bound $F(w(t+1))-F(w^{*})$. We define the error term $e(t)=\nabla F(w(t))-g(t)$
and use dual averaging and some results from \cite{NIPS2011_4247}
to further bound $\mbox{\ensuremath{F(w(t+1))}}-\mbox{\ensuremath{F(w^{*})}}$.
The choice of a nonincreasing sequence of the learning rates $\big\{\alpha(t)\big\}$
for $t\in[T]$ helps achieve the optimal regret and optimality gap
bounds.

\subsection{\label{subsec:Discussion}Discussion}

Our results in Theorem \ref{thm:reg_bnd} and Corollary \ref{cor:opt_gap}
can be interpreted as follows. As long as gradient staleness $\tau\le\mathcal{O}\big(m^{1/4}\big),$
then the expected regret $\mathbb{E}[R(T)]\le\mathcal{O}\big(\sqrt{m}\big)$
and the convergence rate $\mathbb{E}\big[G(T)\big]\le\mathcal{O}\big(1/\sqrt{m}\big)$.
These are the optimal achievable bounds for each metric. We highlight
that in AMB-DG, we can control the gradient staleness by choosing
the appropriate fixed computation time, $T_{p}$. Since in AMB-DG
$\tau=\lceil T_{c}/T_{p}\rceil$,  if we choose $T_{p}$ such that
$\lceil T_{c}/T_{p}\rceil\le\mathcal{O}\big(m^{1/4}\big)$, we can
guarantee $\tau\le\mathcal{O}(m^{1/4})$. Unlike other existing schemes
such as those in \cite{Langford:2009:SLF,hogwild}, the gradient staleness
does not depend on the number of workers. It only depends on the ratio
of $T_{c}$ to $T_{p}$. Hence, AMB-DG can scale to larger cluster
sizes without experiencing the performance degradation that can result
from larger staleness due to larger cluster.

The regret bound in (\ref{eq:regret_bound}) and the optimality gap
bound in (\ref{eq:opt_gap}) depend on $\bar{b}$, $\hat{b}$ and
the ratio $\bar{b}/\hat{b}$. In a real system, these parameters would
be determined by the characteristics of the workers and their computing
speeds. They further depend on the fixed processing time $T_{p}$.
As might be expected, in our experiments on a real distributed system
we observe that $\bar{b}$ and $\hat{b}$ scale linearly with $T_{p}$.
In addition, we observe that the ratio $\bar{b}/\hat{b}$ is typically
bounded by some small constant. Therefore, both bounds in (\ref{eq:regret_bound})
and (\ref{eq:opt_gap}) are asymptotically dominated by $m$.

Since $\hat{b}\le\bar{b}$ with equality if and only if $b(t)=\bar{b}$
for all $t\in[T]$, the regret bound is minimized when the system
observes the same number of data points in each epoch. This is achieved
if there are no stragglers. Hence, the analysis reveals the penalty
incurred due to stragglers. However, this penalty becomes negligible
asymptotically in $m$. Similarly, the dominant term in the convergence
rate bound in (\ref{eq:opt_gap}) is scaled by $\bar{b}/\hat{b}\ge1$.
When there are no stragglers, that dominant term is minimized. Hence,
the ratio $\bar{b}/\hat{b}$ reflects the impact the stragglers can
have on the convergence rate. This impact diminishes as $m$ gets
larger.

\section{Extension to the Fully Decentralized Setting}

The previous analysis assume that the time required by the master
to aggregate gradients and update parameters is negligible. In practice,
the time spent by the master to aggregate gradients increases with
the number of the workers. The added delay can be accounted for an
increase in the effective communication time that each worker node
experiences. For this reason it can be difficult to scale the hub-and-spoke
model to large networks of nodes.

In contrast, the fully decentralized setting, in which no master node
exists, scales well to large network size avoiding significant increases
in communication time. In this setting, each node communicates with
its neighboring nodes using distributed averaging protocols \cite{gossip_boyd}.
During communication, workers exchange information (e.g., primal and
dual variables) with their neighbors a few times and calculate a weighted-average
of the received information to achieve consensus. If perfect consensus
is achieved, then each worker ends up with the same value which corresponds
to the average of the information prior to communication across all
workers. In theory, achieving perfect consensus may require infinite
number of information exchange. Since this is not possible in practice
as workers exchange information a few times, workers often end up
with imperfect consensus.

The model we analyze to investigate the decentralized realization
of AMB-DG consists of two-way noiseless communication channel between
certain pairs of workers. We represent these connections as a graph
${\cal G}(E,V)$ in which each worker $i\in V$ is a vertex in the
graph. Each edge represented two neighbour nodes, i.e., if workers
$i$ and $j$ can communicate with each other then $(i,j)\in E$.
Furthermore, the graph is undirected since if worker $i$ can send
to worker $j$, then also worker $j$ can send to worker $i$. This
means that edge $(i,j)$ is the same as the edge $(j,i)$. We assume
that communication time between all connected pairs is equal. Let
$Q$ be a matrix that is positive semi-definite and doubly stochastic
(i.e., all columns and rows sum to $1$) such that $Q_{ij}>0$ if
$i=j$ or $(i,j)\in E$, otherwise $Q_{ij}=0$. The entries of $Q$
are the weights that workers use to scale the messages received from
their neighbours so that nodes can eventually calculate the average
of their messages. $Q$ is sometimes referred to as communication
matrix.

\subsection{AMB-DG Algorithm in the Masterless Setting}

The version of the decentralized AMB-DG we analyze is similar to the
one explained in Sec. \ref{sec:AMBDG-Scheme}. The main difference
is that instead of master-worker communication, we have inter-node
communication using a distributed averaging protocol. This is similar
to the model we analyzed for the AMB scheme in \cite{amb_iclr}.

\subsubsection*{Gradient Computation}

This step is the same as in the hub-and-spoke setting. Each worker
computes gradients for $T_{p}$ seconds. Due to imperfect consensus,
in each epoch, workers calculate gradients with respect to different
optimization parameters. Let $w_{i}(t)$ be the optimization parameter
at worker $i$ in the $t$-th epoch. The resulting sum of the gradients
computed by worker $i$ in epoch $t$ is

\begin{equation}
g_{i}(t)=\frac{1}{b_{i}(t)}\sum_{s=1}^{b_{i}(t)}\nabla_{w}f\big(w_{i}(t),x_{i}(t,s)\big).
\end{equation}

\subsubsection*{Consensus Phase}

The consensus phase starts immediately after the gradient computation
epoch. At the end of this phase, each worker updates its own optimization
parameter. In each consensus phase, workers spend $T_{c}$ seconds
to exchange information over $r$ rounds of consensus. In the $t$-th
phase, worker $i$ begins by sending message $m_{i}^{(0)}(t)=nb_{i}(t)(z_{i}(t)+g_{i}(t))$.
After $k$ rounds of consensus, where $k\in[r]$, worker $i$ has
the message

\begin{equation}
m_{i}^{(k)}(t)=\sum_{j=1}^{n}Q_{ij}m_{j}^{(k-1)}(t)=\sum_{j=1}^{n}(Q_{ij})^{k}m_{i}^{(0)}(t).
\end{equation}

For a large enough number of iterations, $r$, and if the graph is
connected and the second largest eigenvalue of $Q$ is strictly less
than one, then the consensus rounds converge to $m_{i}^{(r)}(t)=b(t)[\bar{z}(t)+g(t)]$,
where $\bar{z}(t):=\frac{1}{b(t)}\sum_{i=1}^{n}b_{i}(t)z_{i}(t)$,
$g(t)=\frac{1}{b(t)}\sum_{i=1}^{n}g_{i}(t)$. We note that if the
number of iteration per consensus round increases, then the duration
of each consensus phase becomes larger. In other words, as $r$ increases,
$T_{c}$ increases. This means that if $r$ is too large, then gradient
staleness, $\tau$, also increases since $\tau$ increases with $T_{c}$.
We highlight that in the master-worker setup the communication phases
can overlap as can be seen in Fig.\ref{fig:AMBDG-system-model}. In
the fully distributed setting, we also allow such an overlap. To manage
multiple messages, each worker $i$ sends the index of the epoch,
$t$, to its neighbours when sending message $m_{i}^{(r)}(t)$.

\subsubsection*{Parameters Update}

Worker $i$ updates its dual variable after $r$ rounds of consensus
according to

\begin{equation}
z_{i}(t+1)=\frac{1}{b(t)}m_{i}^{(r)}(t)=\bar{z_{i}}(t)+g(t).
\end{equation}

However, often having a large number of consensus rounds may not be
practical and workers complete a small number of rounds. Therefore,
the actual dual variable update contains an additive error term and
can be expressed as

\begin{equation}
z_{i}^{(r)}(t+1)=\bar{z_{i}}(t)+g(t)+\epsilon_{i}(t)
\end{equation}

\noindent where the superscript $r$ in $z_{i}^{(r)}$ denotes the
number of consensus rounds and $\epsilon_{i}(t)$ is the consensus
error at worker $i$. The optimization parameter $w_{i}(t)$ is updated
according to (\ref{eq:primal_var-1}). Let $z(t)$ denote the dual
variable with perfect consensus. According to Lemma 1 from our previous
work \cite{amb_iclr}, let $\delta\geq0$ and $\lambda_{2}(Q)$ be
the second largest eigenvalue of $Q$, then

\begin{equation}
\lVert z_{i}^{(r)}(t+1)-z(t+1)\rVert\le\delta\label{eq:dual_var_bnd}
\end{equation}

\noindent if the number of consensus rounds per worker in each consensus
phase satisfies

\begin{equation}
r\ge\left\lceil \frac{\log(2\sqrt{n}(1+2J/\delta)}{1-\lambda_{2}(Q)}\right\rceil 
\end{equation}

\noindent where $J$ is the Lipschitz parameter of the objective function
$F(w)$.

\subsection{Convergence Results}

The convergence analysis for the fully distributed case is similar
to the one we developed in Sec. \ref{sec:ConvRes} for the hub-and-spoke.
The only exception is that we must account for the consensus error.
In addition, the optimization parameter $w(t)$ in the regret in (\ref{eq:cond_regret})
is replaced with $w_{i}(t)$.
\begin{thm}
\label{thm:regret_decentral} Assume that the workers process gradients
over $T$ epochs during which the expected number of data points workers
sample from $P$ in an i.i.d. manner is $m$. Let $\hat{b}\ne0$ and
$\bar{b}$ be as defined in Sec. \ref{sec:ConvRes}. Choosing $\alpha(t)^{-1}=L+\sqrt{(t+\tau)/\bar{b}}$,
the expected regret for AMB-DG in the fully decentralized setting
with deterministic common gradient delay $\tau$ for all links in
each epoch is

\begin{align}
\mathbb{E}[R(T)]\leq\frac{\bar{b}}{\alpha(T+1)}\psi(w^{*})+2\tau JC\bar{b}\,+\nonumber \\
2LJ^{2}(\tau+1)^{2}\big(1+\log T\big)\bar{b}^{2}+\sigma^{2}\frac{\bar{b}}{\hat{b}}\sqrt{m} & +2J\delta\bar{b}^{3/2}\sqrt{m}.\label{eq:regretFullDec}
\end{align}
\end{thm}
The proof of Theorem \ref{thm:regret_decentral} is provided in Appendix
\ref{app:proof_thm_decent}. The regret bound in (\ref{eq:regretFullDec})
is similar to the one obtained in (\ref{eq:regret_bound}). The only
difference is that (\ref{eq:regretFullDec}) has an extra term related
to the consensus error (i.e., the term with $\delta$). If workers
achieve perfect consensus, then $\delta=0$ and thus (\ref{eq:regretFullDec})
reduces to (\ref{eq:regret_bound}). In contrast, if the communication
time per consensus iteration is long and workers are forced to keep
$r$ small, then $\delta$ will be large and hence the regret bound
is larger.

\section{Numerical Results\label{sec:NumRes}}

In this section we demonstrate the performance of AMB-DG and compare
it to other schemes. We evaluate the performance in two ways. First,
we solve a linear regression problem using a synthetic dataset. We
model the per-node compute time using the shifted exponential distribution.
We compare the performance of AMB-DG with that of AMB under the assumption
of long communication delays. Furthermore, we compare the performance
of AMB-DG with a fixed minibatch delayed-gradient-based scheme called
$K$-batch async \cite{2018arXiv180301113D,Lian:2015:APS:nonconvex}.
$K$-batch async is a suitable baseline to compare AMB-DG with since
both schemes ensure that workers never idle. Hence, they both use
stale gradients. Moreover, both schemes exploit all jobs completed
by all workers. In $K$-batch async, in each epoch each worker calculates
$\tilde{b}/K$ gradients and submits their sum to the master. Here
$\tilde{b}$ is the per-epoch fixed minibatch size and $K$ is an
integer that divides $\tilde{b}$. The master updates the parameter
$w$ after receiving $K$ messages from the workers. These $K$ messages
need not be from distinct workers. Faster workers can contribute more
messages. Second, we compare the performance of AMB-DG with that of
$K$-batch async when training a neural network to classify images
from the CIFAR-10 dataset \cite{krizhevsky2014cifar}. We run these
neural network experiments on SciNet which is an academic high performance
computing cluster. All experiments are repeated $10$ times and the
average performance is reported. We note that the purpose of these
experiments is to compare the performance of our approach that is
based on fixing the compute time with the that of the state-of-the-art
approach that uses a fixed minibatch per epoch. Methods that speed
up convergence when using delayed gradients such as those found in
\cite{AsyncDelayComp} and \cite{Zhang:2016:SAD:3060832.3060950}
can also be applied to AMB-DG.

\subsection{Linear regression}

\subsubsection{Dataset}

For the linear regression problem we generate the global optimizer
vector $w^{*}\in\mathbb{R}^{d}$ from the multivariate Gaussian ${\cal N}({\bf 0,{\bf I})}$.
In each compute epoch, $t$, worker $i$ receives a stream of a sequence
of inputs of the form $(\zeta(t,s),y_{i}(t,s))$. If the $i$-th node
observes $b_{i}(t)$ data points in epoch $t$ then $1\leq s\leq b_{i}(t)$.
The data points $\zeta_{i}(t,s),y_{i}(t,s)$ are generated as following.
First, we generate $\xi_{i}(t,s)\in\mathbb{R}^{d}$ in an i.i.d. manner
according to Gaussian ${\cal N}(\mathbf{0},\mathbf{I})$. Then the
labels $y_{i}(t,s)$ are defined as $y_{i}(t,s)=\zeta_{i}(t,s)^{T}w^{*}+\epsilon_{i}(t,s)$,
where $T$ denotes the transpose operation and the $\epsilon_{i}(t,s)$
are i.i.d. zero-mean Gaussian noise samples with variance $\sigma^{2}$.
In our simulations, we choose $d=10^{4}$ and $\sigma^{2}=10^{-3}$.

\subsubsection{Objective function}

The objective function the system is designed to minimize is

\begin{equation}
F(w)=\sum_{t}\sum_{i=1}^{n}\sum_{s=1}^{b_{i}(t)}\big[\zeta_{i}(t,s)^{T}w-y_{i}(t,s)\big]^{2}.\label{eq:linReg_ObjFunc}
\end{equation}

\noindent We assume that the system initializes the optimization variable
to the all-zero vector $w(1)=0$. In each worker's compute epoch $t$,
the worker calculates gradients $g_{i}(t)$ according to

\begin{equation}
g_{i}(t)=\sum_{s=1}^{b_{i}(t)}\Big[\zeta_{i}(t,s)^{T}w(t)-y_{i}(t,s)\Big]\zeta_{i}(t,s).
\end{equation}

The error rate achieved in the $t$-th epoch is calculated as

\begin{equation}
{\rm Err}(t)=\frac{\lVert A[w(t)-w^{*}]\rVert^{2}}{\lVert Aw^{*}\rVert^{2}}
\end{equation}

\noindent where $A$ is an $N\times d$ matrix whose rows are i.i.d.
${\cal N}(\mathbf{0},\mathbf{I})$ and $\lVert.\rVert$ is the $l_{2}$
norm. We choose $N=250000$. Since $w(1)=0$, the initial error rate
is always ${\rm Err}(1)=1$.

\subsubsection{Shifted exponential model}

To evaluate performance versus wall clock time, we model the time
that each node takes to calculate a fixed local minibatch of size
$b$ as random variable that follows a shifted exponential distribution.
Exponential and shifted exponential distributions are widely used
in the literature to model variable compute times \cite{2018arXiv180301113D,berkeley_speedup_ml}.
We use $T_{i}(t)$ to denote the time taken by worker $i$ to calculate
$b$ gradients in the $t$-th epoch. The probability density function
of $T_{i}(t)$ is

\begin{equation}
f_{T_{i}}(\tau)=\lambda e^{-\lambda(\tau-\xi)},\,\,\tau\ge\xi\ge0\label{eq:shiftExp}
\end{equation}

\noindent where $\xi$ is the minimum duration a single worker takes
to calculates $b$ gradients while $\lambda^{-1}+\xi$ is the expected
time to calculate $b$ gradients.

Conditioned on $T_{i}(t)$, the worker is assumed to make linear progress
through the dataset, i.e., it takes $kT_{i}(t)/b$ to calculate $k$
gradients (note that $k\ge b$ is allowed). In our simulations, we
choose $\lambda=2/3$ and $\xi=1$. Per our model, in both AMB and
AMB-DG, node $i$ is given a fixed time $T_{p}$ to compute $b_{i}(t)$
gradients in epoch $t$. Per our assumptions above, $b_{i}(t)=bT_{p}/T_{i}(t)$,
for some fixed $b\le\mathbb{E}[b_{i}(t)]$. In our simulations, we
assume a network of $n=10$ workers and that $T_{p}=2.5$. We choose
$b=60$ to ensure that the expected minibatch per epoch is $\mathbb{E}[b(t)]\ge nb=600$.
This lower bound is according to Theorem 7 of \cite{amb_iclr}. Moreover,
in our examples, we set $T_{c}=10$.

\subsubsection{Performance evaluation of AMB-DG vs. AMB}

\begin{figure}
\subfloat[\label{fig:ambdg_amb_epoch}Per-epoch performance.]{\includegraphics[scale=0.4]{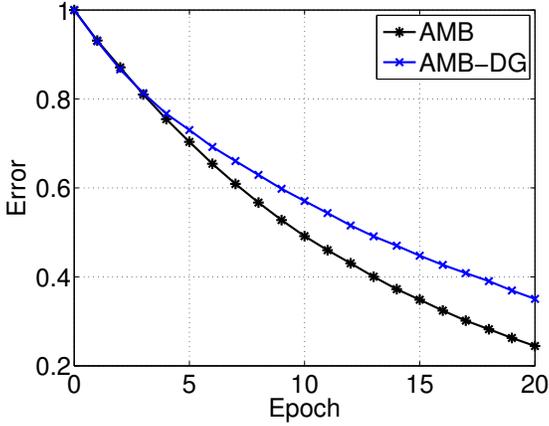}}\hfill{}\subfloat[\label{fig:ambdg_amb_wct}Wall clock time performance.]{\includegraphics[scale=0.4]{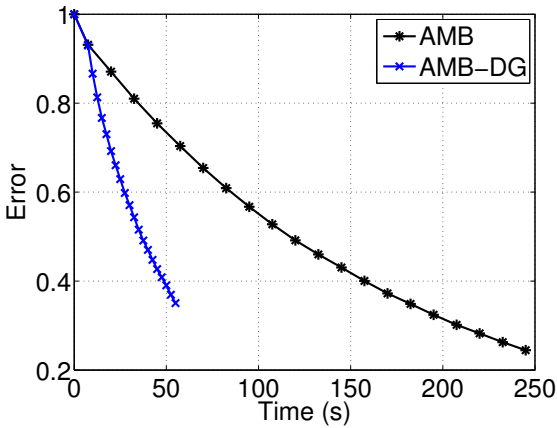}

}

\caption{\label{fig:ambdg_vs_amb}Performance of AMB versus AMB-DG.}

\end{figure}

Figure \ref{fig:ambdg_vs_amb} compares the performance of AMB-DG
with that of AMB. The two plots in Fig. \ref{fig:ambdg_vs_amb} compare
the performance of each scheme with respect to the number of epochs
as well as wall clock time.

Figure \ref{fig:ambdg_amb_epoch} shows that the per-epoch performance
of AMB is better than that of AMB-DG. This is expected since AMB uses
``fresh'' gradients to update the optimization parameter in each
epoch whereas AMB-DG uses stale ones. However, when comparing the
performance against the wall clock time, we observe in Fig. \ref{fig:ambdg_amb_wct}
that AMB-DG is almost three times faster than AMB. In particular,
AMB-DG achieves an error rate of $0.35$ at $55$ ${\rm s}$ whereas
AMB takes about $182$ ${\rm s}$ to achieve the same error rate.
We highlight that (akin to AMB) the gradients in the first epoch of
AMB-DG do not suffer from stalenesss since they are calculated with
respected $w(1)$. Therefore, both schemes achieve similar error rates
after the first epoch. In AMB-DG, gradient staleness increases gradually
until the fifth epoch from which point $\tau=4$. This higher staleness
explains why, as can be observed in Fig.\ref{fig:ambdg_amb_epoch},
AMB-DG achieves higher per-epoch error rates especially from the fifth
epoch onward.

The comparison above highlights one of the shortcomings of AMB. During
communications, AMB workers stay idle. This can slow convergence when
the communication time is long. On the other hand, AMB-DG exploits
the communication time to process additional gradients which, albeit
being stale, help improve convergence speed. The improvement becomes
especially pronounced when communication delays are long. Note that
in AMB, except for the first update, the master updates the optimization
parameters every $T_{p}+T_{c}=12.5$ ${\rm s}$ whereas in AMB-DG
the parameters are updated every $T_{p}=2.5$ ${\rm s}$. The first
update in both schemes takes place after $T_{p}+\frac{1}{2}T_{c}=7.5$
${\rm s}$. Hence, the epoch duration in AMB is larger than that in
AMB-DG. As the communication time is increased, this difference in
epoch durations increases. Conversely, as the communication time approaches
zero, the epoch duration in both schemes become identical. In this
limiting case, the gradient delay in AMB-DG goes to $\tau=0$ which
means that AMB-DG reduces to AMB.

\subsubsection{Performance comparison of AMB-DG with $K$-batch async}

\begin{figure}
\begin{centering}
\includegraphics[scale=0.4]{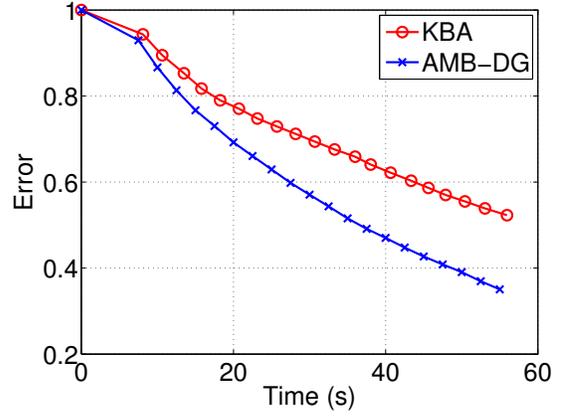}
\par\end{centering}
\caption{\label{fig:AMBDG_vs_KBA_MNIST}AMB-DG vs $K$-batch async.}

\end{figure}

In this experiment we again model the time that takes worker $i$
to calculate $b=60$ gradients in the $t$-th epoch according to the
same shifted exponential distribution in (\ref{eq:shiftExp}). We
set $\lambda=2/3$ and $\xi=1$ as before. We run the simulations
for $n=10$ workers and let $T_{c}=10$. For AMB-DG, we set $T_{p}=2.5$.
Thus, the expected minibatch per epoch in AMB-DG is roughly $600$.
For the $K$-batch async, each worker computes $60$ gradients per
epoch and we set $K=10$. Hence, the master updates the optimization
parameters whenever it receives $\tilde{b}=600$ gradients. Fig. \ref{fig:AMBDG_vs_KBA_MNIST}
illustrates that AMB-DG achieves faster convergence when compared
to $K$-batch async. For instance, the error rate achieved by AMB-DG
after almost $30$ ${\rm s}$ is achieved by $K$-Batch Async at around
$47$ ${\rm s}$ meaning that AMB-DG is over $1.5$ times faster than
K-Batch Async. Observe that the first update takes a long time due
to the long communication delay. This impacts both schemes. If we
ignore this initial shared delay and re-scale time, AMB-DG becomes
about $1.7$ times faster than $K$-Batch Async.

\begin{figure}
\begin{centering}
\includegraphics[scale=0.35]{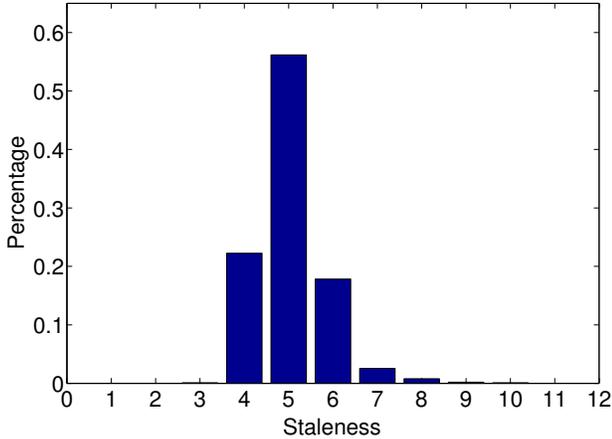}
\par\end{centering}
\caption{\label{fig:stalenessDistKBA}Gradient staleness distribution in $K$-batch
async.}

\end{figure}

To gain further insights into why AMB-DG outperforms $K$-batch async,
we look at the distribution of gradient staleness in both cases. As
explained in Sec. \ref{sec:AMBDG-Scheme}, the gradient staleness
suffered by AMB-DG after the first few epochs is $\tau=T_{c}/T_{p}=4$.
On the other hand, the distribution of gradient staleness experienced
by $K$-batch async as depicted in Figure \ref{fig:stalenessDistKBA}
shows that almost $80\%$ of the gradients received by the master
are delayed by five or more steps. Therefore, although the average
per-epoch minibatch size and the average per-update time in both schemes
are similar, AMB-DG suffers less staleness when compared to $K$-batch
async and thus converges faster.

\subsection{CIFAR-10 Image Classification}

\begin{figure}
\subfloat[\label{fig:cifar10_Tc_10_Err}Training error.]{\begin{centering}
\includegraphics[scale=0.4]{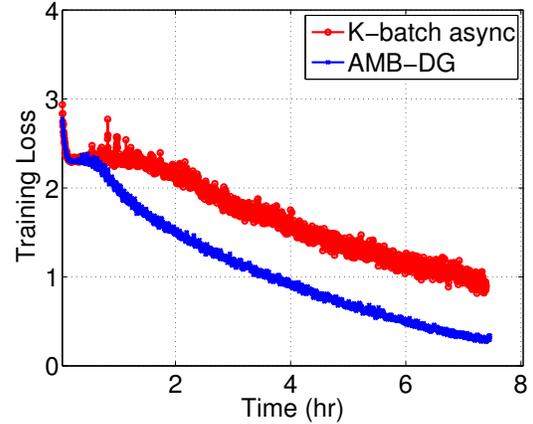}
\par\end{centering}
}\hfill{}\subfloat[\label{fig:cifar10_Tc_10_Acc}Testing accuracy.]{\begin{centering}
\includegraphics[scale=0.4]{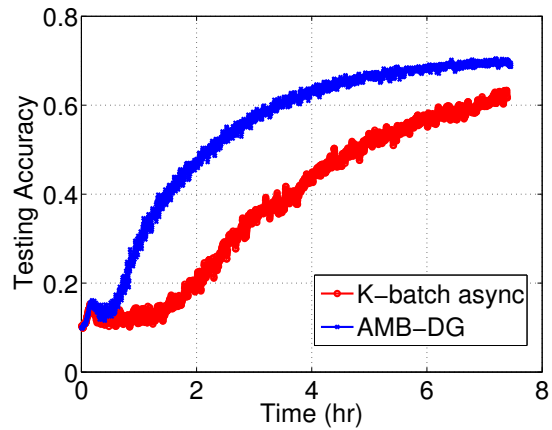}
\par\end{centering}

}

\caption{\label{fig:AMBDG_KBA_CIFAR10_Tc_10}AMB-DG vs. $K$-Batch Async for
CIFAR-10 with induced $T_{c}=10$ sec.}

\end{figure}

Although our analysis assumed convex smooth objective functions, we
complement our work by presenting results for a nonconvex image classification
problem. We train a neural network to classify CIFAR-10 images. We
work with the neural network architecture of \cite{cifar10NN} that
consists of $14$ layers: $9$ convolutional layers followed by $5$
fully connected layers. The loss function that we optimize is the
cross-entropy function. We run our experiments on the SciNet using
$n=4$ workers and a master node. Each node in the SciNet cluster
is equipped with a total of 40 Intel Skylake CPU cores and 202 GB
of memory. Since we use a high performance computing platform, the
inter-node communication time is much smaller than the per-epoch compute
time. Therefore, we artificially delay communication to mimic operating
environments in which communication time is long. We induce $T_{c}=10$
${\rm s}$ while we choose $T_{p}=10$ ${\rm s}$ for AMB-DG. For
$K$-batch async, we set $b=60$ and $K=4$. The per-epoch minibatch
size is in $K$-batch async is $240$ which is approximately the same
as the average minibatch per epoch in AMB-DG. Figure \ref{fig:AMBDG_KBA_CIFAR10_Tc_10}
compares the performance of both schemes when $T_{c}=10$. It illustrates
that AMB-DG achieves a lower training loss and higher testing accuracy
than that achieved by $K$-batch async when the neural network is
trained for the same duration. In particular, the training loss achieved
by AMB-DG after $4$ hours is achieved by $K$-batch async after almost
$7.5$ hours showing that AMB-DG is about $1.9$ times faster than
$K$-batch async. Similarly, the testing accuracy achieved by $K$-batch
async after about $7.5$ hours is already achieved by AMB-DG after
only $4$ hours. We remind the reader that the reported performance
is an average of $10$ experiments and that the purpose of these experiments
is to compare the wall-clock performance of our approach to the existing
fixed minibatch approach. The aim is not to build a deep learning
model that achieves higher training error or testing accuracy than
the existing best in class so, for example, we did not implement dropout
or data augmentation in our simulations.

\subsection{Variable Minibatches of AMB-DG in a Real Distributed System}

\begin{figure}
\begin{centering}
\includegraphics[scale=0.3]{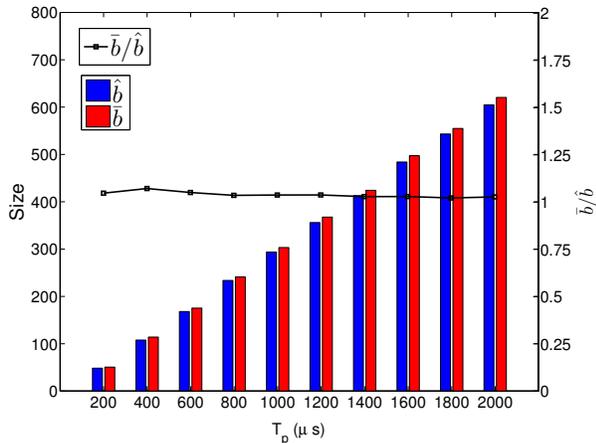}
\par\end{centering}
\caption{\label{fig:bmin_bmax}$\hat{b},\bar{b}$ and $\bar{b}/\hat{b}$ of
AMB-DG.}
\end{figure}

In our analysis in Sec. \ref{sec:ConvRes}, we observe that the regret
depends on several parameters that are system specific. These are
the minimum per-epoch minibatch $\hat{b}$ and the average minibatch
$\bar{b}$. To understand how these parameters scale in real distributed
systems, we ran our algorithm on SciNet using the MNIST dataset \cite{lecun-mnisthandwrittendigit-2010}
for several values of the processing time, $T_{p}$, ranging from
$200$ ${\rm \mu s}$ to $2000$ ${\rm \mu s}$. For each choice of
$T_{p}$, we ran the algorithm for $200$ epochs and recorded the
minimum $\hat{b}$ and the average $\bar{b}$ minibatches across these
$200$ epochs. Fig. \ref{fig:bmin_bmax} shows that these parameters
scale almost linearly with $T_{p}$. Hence, for a given $T_{p}$,
these parameters are bounded and thus the regret bound in (\ref{eq:regret_bound})
and the convergence rate in (\ref{eq:opt_gap}) are dominated by $m$
asymptotically as discussed in Sec. \ref{subsec:Discussion}. Furthermore,
the ratio $\bar{b}/\hat{b}$ is observed to be bounded above by $\gamma<1.1$
as can be read from the right y-axis in Fig. \ref{fig:bmin_bmax}.

\section{Conclusion\label{sec:Conc}}

In this paper, we present the Anytime Minibatch with Delayed Gradients
(AMB-DG) algorithm. AMB-DG is a variable-minibatch scheme for online
distributed stochastic optimization. In AMB-DG, workers are given
a fixed compute time to calculate as many gradients as they can before
submitting the sum of the computed gradients, along with their local
minibatch size, to the master. Following each submission, workers
continue to calculate gradients using the existing optimization parameters,
without waiting for updates from the master. As a result, the master
uses stale gradients to update the optimization parameters. Our theoretical
analysis shows that for smooth convex objective functions, AMB-DG
achieves the optimal regret bound of ${\cal O}(\sqrt{m})$ and the
optimal convergence rate of ${\cal O}(1/\sqrt{m})$, where $m$ is
the total number of samples observed by the entire system and that
the impact of gradient delay is negligible asymptotically in the number
of samples observed, $m$. We compare AMB-DG with AMB under long communication
delay setting when solving a linear regression problem and observe
that AMB-DG outperforms AMB in terms of wall clock time convergence.
We also compare the performance of AMB-DG with that of $K$-batch
async algorithm for the same linear regression problem and observe
that AMB-DG achieves lower staleness and faster convergence in wall
clock time. We also implement AMB-DG and $K$-batch async on a real
distributed system to train a neural network using the CIFAR-10 dataset.
We observe that AMB-DG is almost $2$ times faster.

In our analysis we assumed fixed communication delay. As a future
extension of this work, one can consider a more interesting regime
of operation in which the communication time is random. Furthermore,
this paper analyzed AMB-DG for the master-worker distributed setting.
AMB-DG can be extended to a fully decentralized setup and its performance
and convergence can be analyzed as in \cite{amb_iclr}.

\section{Acknowledgement}

We thank Jason Lam and Zhenhua Hu of Huawei Technologies Canada and
Tharindu Adikari from the University of Toronto for technical discussions.
We thank Compute Canada for providing their high performance computing
platform to run our experiments throughout this project.

\bibliographystyle{IEEEtran}
\bibliography{References}

\appendices{}

\section{AMB-DG Convergence Proofs\label{appendix:Proofs}}

\subsection{Proof of Theorem \ref{thm:reg_bnd}}

The proof of Theorem \ref{thm:reg_bnd} uses the convexity and Lipschitz
smoothness properties of $F(w)$ and $f(w,x)$. We also use the following
three lemmas to prove some intermediate results. Lemma \ref{lem:lemm_x+}
is from \cite{Nesterov:2009:dual_average} and Lemmas \ref{lem:delaybnd}
and \ref{lem:lemma7delaybnd} are from \cite{NIPS2011_4247}.
\begin{lem}
\label{lem:lemm_x+} If $x^{+}=\arg\min_{u\in{\cal X}}\langle z,u\rangle+A\psi(u)$,
then for any $x,z\in\mathcal{X}$
\[
\langle z,x\rangle+A\psi(x)\ge\langle z,x^{+}\rangle+A\psi(x^{+})+AD_{\psi}(x,x^{+}),
\]

\noindent where $D_{\psi}(x,x^{+})$ is the Bregman divergence between
$x$ and $x^{+}$.
\end{lem}
\begin{lem}
\label{lem:delaybnd} Let the assumptions of Sec. \ref{subsec:Preliminaries}
hold. Then for any sequence $w(t)$,

\begin{align*}
\sum_{t=1}^{T}\big\langle\nabla F(w(t))-\nabla F(w(t-\tau)), & w(t+1)-w^{*}\big\rangle\\
\le\frac{L}{2}\sum_{t=1}^{T}\big\lVert w(t-\tau) & -w(t+1)\big\rVert^{2}+2\tau JC.
\end{align*}

Consequently, if $\mathbb{E}\big[\lVert w(t)-w(t+1)\rVert^{2}\big]\le4J^{2}\kappa(t)^{2}$
for a nonincreasing sequence $\kappa(t)$,

\begin{align*}
\mathbb{E}\Bigg[\sum_{t=1}^{T}\big\langle\nabla F(w(t))-\nabla F(w(t-\tau)) & ,w(t+1)-w^{*}\big\rangle\Bigg]\\
\le2LJ^{2}(\tau+1)^{2} & \sum_{t=1}^{T}\kappa(t-\tau)^{2}+2\tau JC.
\end{align*}
\end{lem}
\begin{lem}
\label{lem:lemma7delaybnd}Let the assumptions of Sec.\ref{subsec:Preliminaries}
hold. Let $\alpha(t)^{-1}=L+\eta(t)$ for $\eta(t)>0$ and $L\ge0$.
Then $\mathbb{E}\Big[\lVert w(t)-w(t+1)\rVert^{2}\Big]\le\frac{4J^{2}}{\eta(t)^{2}}$.
\end{lem}
The convergence of our scheme is impacted by the following two factors.
First, gradients are calculated with respect to $f(w,x)$ rather than
$F(w)$. Furthermore, in each epoch $t$, the master receives the
gradients $g(t)$ which are delayed by $\tau$ steps. We define the
error term $e(t)$.

\begin{equation}
e(t):=\nabla F(w(t))-g(t)\label{eq:errTerm}
\end{equation}

First, we look at $F(w(t))-F(w^{*})$ and develop some bounds.

\begin{align}
F(w(t))-F(w^{*}) & \le\big\langle\nabla F(w(t)),w(t)-w^{*}\big\rangle\nonumber \\
=\langle\nabla F(w(t)), & w(t+1)-w^{*}\rangle\label{F_F*_cnvxty_bnd}\\
 & +\langle\nabla F(w(t)),w(t)-w(t+1)\rangle\\
\le\langle\nabla F(w(t)), & w(t+1)-w^{*}\rangle+F(w(t))\nonumber \\
-F & (w(t+1))+\frac{L}{2}\big\lVert w(t)-w(t+1)\big\rVert^{2},\label{F_F*_L_smooth}
\end{align}

\noindent where (\ref{F_F*_cnvxty_bnd}) follows from the convexity
of $F$ and (\ref{F_F*_L_smooth}) from the $L$-Lipschitz continuity
of $\nabla F$, cf., (\ref{eq:smoothProp}).

From (\ref{eq:errTerm}), $\nabla F(w(t))=e(t)+g(t)$. Substituting
this in (\ref{F_F*_L_smooth}) and rearranging, we get a bound for
$F(w(t+1))-F(w^{*})$.

\begin{align}
F(w(t+1))-F(w^{*}) & \le\langle e(t)+g(t),w(t+1)-w^{*}\rangle\nonumber \\
 & +\frac{L}{2}\big\lVert w(t)-w(t+1)\big\rVert^{2}\\
=\langle e(t),w(t+1)- & w^{*}\rangle+\langle g(t),w(t+1)-w^{*}\rangle\nonumber \\
 & +\frac{L}{2}\big\lVert w(t)-w(t+1)\big\rVert^{2}\\
=\langle z(t+1)-z(t), & w(t+1)-w^{*}\rangle\nonumber \\
+\langle e(t),w(t+1) & -w^{*}\rangle+\frac{L}{2}\big\lVert w(t)-w(t+1)\big\rVert^{2}\label{F_F*_sub_z}\\
=\langle z(t+1),w(t+ & 1)-w^{*}\rangle-\langle z(t),w(t+1)-w^{*}\rangle\nonumber \\
+\langle e(t),w(t+1) & -w^{*}\rangle+\frac{L}{2}\big\lVert w(t)-w(t+1)\big\lVert^{2},\label{F_F*_sub_z_2}
\end{align}
where (\ref{F_F*_sub_z}) follows by observing that $g(t)=z(t+1)-z(t)$
from (\ref{eq:dual_var-1}).

Recall that $w(t)=\arg\min_{w\in{\cal W}}\left\{ \langle z(t),w\rangle+\alpha(t)^{-1}\psi(w)\right\} $
according to (\ref{eq:primal_var-1}). We apply Lemma \ref{lem:lemm_x+}
to (\ref{F_F*_sub_z_2}) where we replace $x^{+}$, $x$, $z$ and,
$A$ from Lemma \ref{lem:lemm_x+} with $w(t)$, $w(t+1)$, $z(t)$,
and $\alpha(t)^{-1}$, respectively, and re-arrange to get

\begin{align}
-\langle z(t),w(t+1)-w^{*}\rangle\le & -\langle z(t),w(t)-w^{*}\rangle\nonumber \\
+\frac{1}{\alpha(t)} & \big[\psi(w(t+1))-\psi(w(t))\big]\nonumber \\
 & -\frac{1}{\alpha(t)}D_{\psi}\big(w(t+1),w(t)\big).\label{lemm_x+_dual_avg}
\end{align}
Substitute (\ref{lemm_x+_dual_avg}) in (\ref{F_F*_sub_z_2}) yields

\begin{align}
F(w(t+1))-F(w^{*})\le & \langle z(t+1),w(t+1)-w^{*}\rangle\nonumber \\
-\langle z(t),w(t)-w^{*}\rangle & +\frac{1}{\alpha(t)}\big[\psi(w(t+1))-\psi(w(t))\big]\nonumber \\
+\langle e(t),w(t+1)- & w^{*}\rangle-\frac{1}{\alpha(t)}D_{\psi}\big(w(t+1),w(t)\big)\nonumber \\
 & +\frac{L}{2}\lVert w(t)-w(t+1)\rVert^{2}.\label{F_F*_sub_z_3}
\end{align}
Since $\alpha(t)$ is a sequence of positive nonincreasing step sizes,
define a positive nondecreasing sequence $\eta(t)$ such that $\alpha(t)^{-1}=L+\eta(t)$.
As will be seen later, $\eta(t)$ will subsequently be optimized to
achieve the desired regret bound. Typically, $\eta(t)$ is set proportional
to $\sqrt{t+\tau}$. Substituting $L+\eta(t)$ for $\alpha(t)^{-1}$
and observing that $\frac{1}{2}\lVert w(t)-w(t+1)\rVert^{2}\le D_{\psi}\big(w(t+1),w(t)\big)$
since $\psi(x)$ is strongly convex, and then noticing that the $LD_{\psi}(w(t+1),w(t))$
term cancels. Equation (\ref{F_F*_sub_z_3}) then becomes

\begin{align}
F(w(t+1))-F(w^{*})\leq\nonumber \\
\langle z(t+1),w(t+1)-w^{*} & \rangle-\langle z(t),w(t)-w^{*}\rangle\nonumber \\
+\langle e(t),w(t+1)-w^{*}\rangle & +\frac{1}{\alpha(t)}\big[\psi(w(t+1))-\psi(w(t))\big]\nonumber \\
 & -\eta(t)D_{\psi}\big(w(t+1),w(t)\big).\label{F_F*_sub_z_4}
\end{align}
Now we sum both sides of (\ref{F_F*_sub_z_4}) from $t=1$ to $t=T$
and note that the term $\langle z(1),w(1)-w^{*}\rangle=0$ since $z(1)=0$
to get

\begin{align}
\sum_{t=1}^{T}F\big(w(t+1)\big)-F(w^{*}) & \leq\nonumber \\
\big\langle z(T+1),w(T+1)-w^{*} & \big\rangle+\frac{\psi(w(T+1))}{\alpha(T)}\nonumber \\
-\frac{\psi(w(1))}{\alpha(1)}+\sum_{t=2}^{T}\psi(w(t)) & \Big[\frac{1}{\alpha(t-1)}-\frac{1}{\alpha(t)}\Big]\nonumber \\
-\sum_{t=1}^{T}\eta(t)D_{\psi}\big(w(t+1),w(t & )\big)+\sum_{t=1}^{T}\langle e(t),w(t+1)-w^{*}\rangle\\
\le\big\langle z(T+1),w(T+1)- & w^{*}\big\rangle+\frac{\psi(w(T+1))}{\alpha(T+1)}\nonumber \\
-\sum_{t=1}^{T}\eta(t)D_{\psi}\big(w(t+1),w(t & )\big)+\sum_{t=1}^{T}\langle e(t),w(t+1)-w^{*}\rangle\label{F_F*_sub_z_5_1}\\
\le\frac{\psi(w^{*})}{\alpha(T+1)}-\sum_{t=1}^{T}\eta(t)D_{\psi} & \big(w(t+1),w(t)\big)\nonumber \\
+ & \sum_{t=1}^{T}\langle e(t),w(t+1)-w^{*}\rangle\label{F_F*_sub_z_5}
\end{align}
where (\ref{F_F*_sub_z_5_1}) follows from the facts that $\frac{1}{\alpha(t-1)}-\frac{1}{\alpha(t)}\le0$
(and hence $\alpha(T)\ge\alpha(T+1)$) and $\psi(w)~\ge~0$ so we
can drop the negative term $-\frac{\psi(w(1))}{\alpha(1)}$. Equation
(\ref{F_F*_sub_z_5}) follows from that $\big\langle z(T+1),w(T+1)-w^{*}\rangle+\frac{\psi(w(T+1))}{\alpha(T+1)}\le\frac{\psi(w^{*})}{\alpha(T+1)}$.
To see this, observe that according to dual average rule $w(T+1)=\arg\min_{w\in\mathcal{W}}\Big\{\big\langle z(T+1),w\big\rangle+\frac{1}{\alpha(T+1)}\psi(w)\Big\}$.
So, if we substitute $w^{*}$ for $w(T+1)$ in the first two term
on the right hand side of (\ref{F_F*_sub_z_5_1}), the expression
will be larger. This substitution eliminates the inner product term
and we get (\ref{F_F*_sub_z_5}).

Next, we look at the last term in (\ref{F_F*_sub_z_5}); i.e., $\mbox{\ensuremath{\sum_{t=1}^{T}\langle e(t),w(t+1)-w^{*}\rangle}}$.

\begin{align}
\sum_{t=1}^{T}\langle e(t),w(t+1)-w^{*}\rangle=\nonumber \\
\sum_{t=1}^{T}\langle\nabla F(w(t))-\nabla F(w(t & -\tau)),w(t+1)-w^{*}\rangle\nonumber \\
+\sum_{t=1}^{T}\langle\nabla F(w(t-\tau)) & -g(t),w(t+1)-w^{*}\rangle.\label{sum_err}
\end{align}
Using Lemma \ref{lem:delaybnd}, and noting that from Lemma \ref{lem:lemma7delaybnd}
that $\mathbb{E}\Big[\lVert w(t)-w(t+1)\rVert^{2}\Big]\le\frac{4J^{2}}{\eta(t)^{2}}$,
we can bound the expected value of the first term on the right hand
side of (\ref{sum_err}) as

\begin{align}
\mathbb{E}\Bigg[\sum_{t=1}^{T} & \langle\nabla F(w(t))-\nabla F(w(t-\tau)),w(t+1)-w^{*}\rangle\Bigg]\nonumber \\
 & \leq2\tau JC+2LJ^{2}(\tau+1)^{2}\sum_{t=1}^{T}\frac{1}{\eta(t-\tau)^{2}},\label{err_1st_term_bnd}
\end{align}

\noindent where the nonincreasing sequence $\kappa(t)$ in Lemma \ref{lem:delaybnd}
is replaced by $\frac{1}{\eta(t)}$ since $\eta(t)$ is nondecreasing.

Now, we look at the second term in (\ref{sum_err})

\begin{align}
\langle\nabla F(w(t-\tau))-g(t),w(t+1)-w^{*}\rangle\nonumber \\
=\langle\nabla F(w(t-\tau))-g(t),w(t)-w^{*}\rangle\nonumber \\
+\langle\nabla F(w(t-\tau))-g(t), & w(t+1)-w(t)\rangle\\
\le\langle\nabla F(w(t-\tau))-g(t),w(t)-w^{*}\rangle & +\nonumber \\
\frac{1}{2\eta(t)}\big\lVert\nabla F(w(t-\tau))-g(t)\big\rVert^{2}+\frac{\eta(t)}{2} & \lVert w(t+1)-w(t)\rVert^{2}.\label{err_2nd_term_bnd}
\end{align}
The last inequality follows from the Fenchel-Young inequality which
states that for every $u,v\in\mathbb{R}^{n}$ given a real valued
convex function $h(u)$ and its convex conjugate $h^{*}(u)$, the
inner product $\mbox{\ensuremath{\langle u,v\rangle\le h(u)+h^{*}(v)}}$.
In (\ref{err_2nd_term_bnd}), we let $\mbox{\ensuremath{u=\nabla F(w(t-\tau))-g(t)}}$,
$\mbox{\ensuremath{v=w(t+1)-w(t)}}$, $\mbox{\ensuremath{h(u)=\frac{\eta(t)}{2}\lVert u\rVert^{2}}}$
and $\mbox{\ensuremath{h^{*}(v)=\frac{1}{2\eta(t)}\lVert v\rVert^{2}}}$.
Note that conditioned on $g(1),g(2),...,g(t-1)$, $w(t)$ is not random
and hence $w(t)$ does not depend on $\nabla F(w(t-\tau))-g(t)$.
Furthermore, since by definition $\nabla F(w)=\mathbb{E}\big[\nabla f(w,x)\big]$,
then $\mathbb{E}\big[\nabla F(w(t-\tau))-g(t)\big]=0$. Hence, the
first term on the right hand side of (\ref{err_2nd_term_bnd}) has
zero expectation. Due to the bounded variance assumption, and assuming
that the master receives $b(t)$ gradients in epoch $t$,

\begin{align*}
\mathbb{E}\Big[\big\lVert\nabla F(w(t-\tau))-g(t)\big\rVert^{2}\big|b(t)\Big] & =\\
\mathbb{E}\Big[\big\lVert\nabla F(w(t-\tau))-\frac{1}{b(t)}\sum_{i=1}^{n}\sum_{s=1}^{b_{i}(t)} & \big[f(w(t-\tau),x_{i}(t,s))\big]\big\rVert^{2}\big|b(t)\Big]\\
\leq & \frac{\sigma^{2}}{b(t)}.
\end{align*}

The bound above follows from the fact that $x_{i}(t,s)$ are sampled
i.i.d, so the cross terms have zero expectation. Taking the expected
value of (\ref{err_2nd_term_bnd}) conditioned on $b(t)$

\begin{align}
\mathbb{E}\Bigg[\langle\nabla F(w(t-\tau))-g(t),w(t+1)-w^{*}\rangle\big|b(t)\Bigg]\nonumber \\
\le\frac{1}{2\eta(t)}\frac{\sigma^{2}}{b(t)}+\frac{\eta(t)}{2}\mathbb{E}\Big[\big\lVert w(t+1) & \big\rVert^{2}\Big].\label{err_2nd_term_bnd_exp}
\end{align}
Taking the expectation on both sides in (\ref{sum_err}) conditioned
on ${\cal B}_{{\rm tot}}:=\big\{ b(t),t\in[T]\big\}$ and substituting
(\ref{err_1st_term_bnd}) and (\ref{err_2nd_term_bnd_exp}),

\begin{align}
\sum_{t=1}^{T}\mathbb{E}\Big[\langle e(t),w(t+1)-w^{*}\rangle\big|{\cal B}_{{\rm tot}}\Big] & \le2\tau JC+\nonumber \\
2LJ^{2}(\tau+1)^{2}\sum_{t=1}^{T}\frac{1}{\eta(t-\tau)^{2}} & +\frac{\sigma^{2}}{2}\sum_{t=1}^{T}\frac{1}{\eta(t)b(t)}\nonumber \\
+\sum_{t=1}^{T}\frac{\eta(t)}{2}\mathbb{E}\Big[\lVert w(t+ & 1)-w(t)\rVert^{2}\Big].\label{eq:exp_err_term_bnd}
\end{align}
Substituting this into (\ref{F_F*_sub_z_5}) after taking the conditional
expectation given ${\cal B}_{{\rm tot}}$ and observing that the term
$\mbox{\ensuremath{-\eta(t)D_{\psi}(w(t+1),w(t))}}$ in (\ref{F_F*_sub_z_5})
is upper bounded by $\mbox{\ensuremath{-\frac{\eta(t)}{2}\lVert w(t+1)-w(t)\rVert^{2}}}$
since $\mbox{\ensuremath{\frac{1}{2}\lVert w(t+1)-w(t)\rVert^{2}\le D_{\psi}(w(t+1),w(t))}}$.
This then cancels the last term in (\ref{eq:exp_err_term_bnd}), and
we have

\begin{align}
\mathbb{E}\Big[\sum_{t=1}^{T}F\big(w(t+1)\big)-F(w^{*})\Big]\nonumber \\
=\mathbb{E}\bigg[\mathbb{E}\Big[\sum_{t=1}^{T}F\big(w(t+1)\big)-F & (w^{*})\big|{\cal B}_{{\rm tot}}\Big]\bigg]\\
\le\frac{1}{\alpha(T+1)}\psi(w^{*})+2\tau JC\nonumber \\
+2LJ^{2}(\tau+1)^{2}\sum_{t=1}^{T}\frac{1}{\eta(t-\tau)^{2}} & +\frac{\sigma^{2}}{2}\mathbb{E}\bigg[\sum_{t=1}^{T}\frac{1}{\eta(t)b(t)}\bigg].\label{Expct_F_F*}
\end{align}
Let $\eta(t)=\sqrt{(t+\tau)/\bar{b}}$ and recalling that, $\mathbb{E}[1/b(t)]\le1/\hat{b}$
and $m=T\bar{b}$, then
\begin{align}
\sum_{t=1}^{T}\frac{1}{\eta(t)\hat{b}}=\sum_{t=1}^{T}\frac{\sqrt{\bar{b}}}{\hat{b}\sqrt{t+\tau}} & \le\frac{\sqrt{\bar{b}}}{\hat{b}}\sum_{t=1}^{T}\frac{1}{\sqrt{t+1}}\nonumber \\
\le & 2\frac{\sqrt{\bar{b}}\sqrt{T}}{\hat{b}}=\frac{2\sqrt{m}}{\hat{b}},\label{sumEta}
\end{align}
and,

\begin{equation}
\sum_{t=1}^{T}\frac{\bar{b}}{\eta(t-\tau)^{2}}=\bar{b}\sum_{t=1}^{T}\frac{1}{t}\le\bar{b}\big[1+\log{\big(T\big)}\big].\label{sumEtaTau}
\end{equation}

The expected regret after $T$ epochs is

\begin{align*}
\mathbb{E}\big[R(T)\big]=\mathbb{E}\bigg[\sum_{t=1}^{T}\sum_{i=1}^{n}\sum_{s=1}^{b_{i}(t)}f(w & (t+1),x_{i}(t+1,s))\\
- & f(w^{*},x_{i}(t+1,s))\bigg]\\
=\sum_{t=1}^{T}\sum_{i=1}^{n}\mathbb{E}\bigg[\sum_{s=1}^{b_{i}(t)}f(w(t+1),x_{i} & (t+1,s))\\
- & f(w^{*},x_{i}(t+1,s))\bigg]\\
=\sum_{t=1}^{T}\sum_{i=1}^{n}\mathbb{E}\bigg[\mathbb{E}\bigg[\sum_{s=1}^{b_{i}(t)}f(w(t+1) & ,x_{i}(t+1,s))\\
-f(w^{*} & ,x_{i}(t+1,s))\big\vert b_{i}(t)\bigg]\bigg]
\end{align*}

\begin{align}
=\sum_{t=1}^{T}\sum_{i=1}^{n}\mathbb{E}\bigg[\sum_{s=1}^{b_{i}(t)}\mathbb{E}\Big[F(w(t+1) & -F(w^{*})\Big]\bigg]\label{eq:EF_Ef}\\
=\sum_{t=1}^{T}\sum_{i=1}^{n}\mathbb{E}\Big[b_{i}(t)\mathbb{E}\Big[F(w(t+1) & -F(w^{*})\Big]\Big]\nonumber \\
=\sum_{t=1}^{T}\sum_{i=1}^{n}\mathbb{E}\Big[b_{i}(t)\Big]\mathbb{E}\Big[F(w(t+1 & )-F(w^{*})\Big]\label{eq:b_F_indep}\\
\le\sum_{t=1}^{T}\sum_{i=1}^{n}\frac{\bar{b}}{n}\mathbb{E}\Big[F(w(t+1)-F & (w^{*})\Big]\label{eq:regret_bnd_1}
\end{align}

\noindent where equation (\ref{eq:EF_Ef}) follows from the fact that
$\mathbb{E}[f(w(t),x(t))]=\mathbb{E}[\mathbb{E}[f(w(t),x(t))\lvert w(t)]]=\mathbb{E}[F(w(t))]$.

Substituting (\ref{sumEta}), (\ref{sumEtaTau}), and (\ref{Expct_F_F*})
in (\ref{eq:regret_bnd_1}), we get

\begin{align}
\mathbb{E}\big[R(T)\big]\le & \frac{\bar{b}}{\alpha(T+1)}\psi(w^{*})+2\tau JC\bar{b}\,+\nonumber \\
 & 2LJ^{2}(\tau+1)^{2}\big(1+\log T\big)\bar{b}^{2}+\sigma^{2}\frac{\bar{b}}{\hat{b}}\sqrt{m}.\label{Expct_F_F*_2}
\end{align}

Finally, the bound in (\ref{eq:regret_bound}) of Theorem \ref{thm:reg_bnd}
is obtained from (\ref{Expct_F_F*_2}) by observing that $\psi(w^{*})\le C^{2}/2$
and $\alpha(t)^{-1}=L+\sqrt{(t+\tau)/\bar{b}}$. Hence, $1/\alpha(T+1)=L+\sqrt{(T+1+\tau)/\bar{b}}$
since $\bar{b}=m/T$.

\subsection{Proof of Corollary \ref{cor:opt_gap}}

Let $\hat{w}(T)=\frac{1}{T}\sum_{t=1}^{T}w(t+1)$, then the expected
optimality gap can be bounded by

\begin{align}
\mathbb{E}\big[G(T)\big] & =\mathbb{E}\big[F(\hat{w}(T))\big]-F(w)\\
=\mathbb{E}\Bigg[F\bigg( & \frac{1}{T}\sum_{t=1}^{T}w(t+1)\bigg)\Bigg]-F(w^{*})\\
\le\mathbb{E}\Bigg[\frac{1}{T} & \sum_{t=1}^{T}F\big(w(t+1)\big)\Bigg]-F(w^{*})\label{optGapPf_1}\\
=\mathbb{E}\Bigg[\frac{\bar{b}}{m} & \sum_{t=1}^{T}\Big(F\big(w(t+1)\big)-F(w^{*})\Big)\Bigg]\\
=\frac{\bar{b}}{m}\mathbb{E}\Bigg[ & \sum_{t=1}^{T}\Big(F\big(w(t+1)\big)-F(w^{*})\Big)\Bigg]\\
\le\,\,\,\bar{b}\Bigg( & \frac{C^{2}}{2m}\Big(L+\sqrt{(T+1+\tau)/\bar{b}}\Big)+\frac{2\tau JC}{m}+\nonumber \\
 & \frac{2LJ^{2}(\tau+1)^{2}\bar{b}\big(1+\log T\big)}{m}+\frac{\sigma^{2}}{\hat{b}\sqrt{m}}\Bigg).\label{Expct_F_F*_3}
\end{align}

\noindent where (\ref{optGapPf_1}) follows from the convexity of
$F(w)$ and (\ref{Expct_F_F*_3}) follows from Theorem \ref{thm:reg_bnd}.

\section{Proof of Theorem \ref{thm:regret_decentral}\label{app:proof_thm_decent}}

Define the sample path ${\cal B}_{{\rm tot}}=\{b_{i}(t),i\in[n],t\in[T]\}$.
Then conditioned on this sample path, the expected regret is given
by

\begin{align}
\mathbb{E}\big[R(T)\lvert{\cal B_{{\rm tot}}}\big]=\nonumber \\
\mathbb{E}\bigg[\sum_{t=1}^{T}\sum_{i=1}^{n}\sum_{s=1}^{b_{i}(t)}f( & w_{i}(t+1),x_{i}(t+1,s))\nonumber \\
 & -f(w^{*},x_{i}(t+1,s))\Big\lvert{\cal B}_{{\rm tot}}\bigg]\\
=\sum_{t=1}^{T}\sum_{i=1}^{n}\mathbb{E}\bigg[\sum_{s=1}^{b_{i}(t)} & F(w_{i}(t+1))-F(w^{*})\bigg]\\
=\sum_{t=1}^{T}\sum_{i=1}^{n}\mathbb{E}\bigg[\sum_{s=1}^{b_{i}(t)} & F(w(t+1))-F(w^{*})\nonumber \\
 & +F(w_{i}(t+1))-F(w(t+1)\bigg]\\
=\sum_{t=1}^{T}\sum_{i=1}^{n}\mathbb{E}\bigg[\sum_{s=1}^{b_{i}(t)} & F(w(t+1))-F(w^{*})\bigg]+\nonumber \\
\sum_{t=1}^{T}\sum_{i=1}^{n}\sum_{s=1}^{b_{i}(t)} & \mathbb{E}\Big[F(w_{i}(t+1))-F(w(t+1))\Big]\\
\leq\sum_{t=1}^{T}\sum_{i=1}^{n}\mathbb{E}\bigg[\sum_{s=1}^{b_{i}(t)} & F(w(t+1))-F(w^{*})\bigg]+\nonumber \\
\sum_{t=1}^{T}\sum_{i=1}^{n} & \sum_{s=1}^{b_{i}(t)}J\mathbb{E}\Big[\lVert w_{i}(t+1)-w(t+1)\rVert\Big]\label{eq:bnd_lipschitz}
\end{align}

\noindent where (\ref{eq:bnd_lipschitz}) follows from the Lipschitz
continuity assumption (\ref{eq:G-Lipschitz}). To bound the the term
$\lVert w_{i}(t+1)-w(t+1)\rVert$ above we use Lemma 2 from \cite{Tsianos2016EfficientDO}
which states that for all $i\in[n]$ and all $t\in[T],$

\begin{equation}
\lVert w_{i}(t+1)-w(t+1)\rVert\le\alpha(t+1)\lVert z_{i}^{(r)}(t+1)-z(t+1)\rVert.\label{eq:Lem2Tsianos_bnd}
\end{equation}

Substituting (\ref{eq:dual_var_bnd}) and (\ref{eq:Lem2Tsianos_bnd})
in (\ref{eq:bnd_lipschitz}), we get

\begin{align}
\mathbb{E}\big[R(T)\lvert{\cal B_{{\rm tot}}}\big]\leq\sum_{t=1}^{T}\sum_{i=1}^{n}\mathbb{E}\bigg[\sum_{s=1}^{b_{i}(t)} & F(w(t+1))-F(w^{*})\bigg]\nonumber \\
+\sum_{t=1}^{T} & \sum_{i=1}^{n}\sum_{s=1}^{b_{i}(t)}J\delta\alpha(t+1)\\
=\sum_{t=1}^{T}\sum_{i=1}^{n}\mathbb{E}\bigg[\sum_{s=1}^{b_{i}(t)} & F(w(t+1))-F(w^{*})\bigg]\nonumber \\
+ & \sum_{t=1}^{T}J\delta b(t)\alpha(t+1).
\end{align}

Since $\mathbb{E}\big[R(T)\big]=\mathbb{E}\big[\mathbb{E}\big[R(T)\lvert{\cal B_{{\rm tot}}}\big]\big]$,
then the expected regret is

\begin{align}
\mathbb{E}\big[R(T)\big]\leq\sum_{t=1}^{T}\sum_{i=1}^{n}\mathbb{E}\bigg[\mathbb{E}\bigg[\sum_{s=1}^{b_{i}(t)} & F(w(t+1))-F(w^{*})\bigg]\bigg]\nonumber \\
+ & J\delta\bar{b}\sum_{t=1}^{T}\alpha(t+1).\label{eq:expct_regret_dec_2terms}
\end{align}

We look at the second term $\sum_{t=1}^{T}J\delta\bar{b}\alpha(t+1)$
and observe that $\alpha(t+1)\le1/\sqrt{(t+\tau)/\bar{b}}$. Thus,

\begin{align}
J\delta\bar{b}\sum_{t=1}^{T}\alpha(t+1)\le & J\delta\bar{b}\sum_{t=1}^{T}\frac{1}{\sqrt{(t+\tau)/\bar{b}}}\\
= & j\delta\bar{b}^{3/2}\sum_{t=1}^{T}\frac{1}{\sqrt{(t+\tau)}}\\
\leq & 2J\delta\bar{b}^{3/2}\sqrt{m}.
\end{align}

The first term in (\ref{eq:expct_regret_dec_2terms}) is the expected
regret for the hub-and-spoke case so we can substitute its bound from
(\ref{Expct_F_F*_2}). Thus, we have

\begin{align}
\mathbb{E}\big[R(T)\big]\leq\frac{\bar{b}}{\alpha(T+1)} & \psi(w^{*})+2\tau JC\bar{b}\,+\nonumber \\
2LJ^{2}(\tau+1)^{2}\big(1+ & \log T\big)\bar{b}^{2}+\sigma^{2}\frac{\bar{b}}{\hat{b}}\sqrt{m}+2J\delta\bar{b}^{3/2}\sqrt{m}.
\end{align}

\section{AMB-DG Pseudocode\label{app:AMB-DG-Pseudocode}}

In this section, we present pseudocode for AMB-DG. Algorithm \ref{alg:ambdg_worker}
details the process at the workers while Algorithm \ref{alg:ambdg_master}
shows the AMB-DG steps at the master. In Algorithm \ref{alg:ambdg_worker},
line 2 is about parameter initialization at the workers. Lines 4-11
correspond to the gradient compute phase. Line 12 is related to worker-to-master
communication while in lines 8-10 and 13-15, workers receive the update
from the master and update their local versions of the optimization
parameters. On the other hand, in Algorithm \ref{alg:ambdg_master},
lines 3-7 are related to worker-to-master communication in which the
master receives messages from the workers. Lines 8-10, correspond
to the parameter update phase while line 12 represent the communication
from the master to all workers.

\begin{algorithm}
\begin{algorithmic}[1]  	
\ForAll {$t = 1, 2, ...$}
		\State initialize $g_i(t) = 0, b_i(t) = 0, s = 0$
		\State $T_{0} = \texttt{current\_time}$
		\While{$\texttt{current\_time} \le T_p - T_{0}$}
			\State sample i.i.d input data $x_{i}(t,s)$ from $P$
			\State calculate $g_i(t) = g_i(t) + \nabla f(w(t), x_{i}(t,s))$
			\State $b_i(t)$++, $s$++.
		\EndWhile
		\State send $g_i(t)$, $b_i(t)$ to the master
		\If {received updated parameter $\tilde{w}$ from master}
				\State set $w(t)=\tilde{w}$
		\EndIf
	\EndFor 
\end{algorithmic}

\caption{\label{alg:ambdg_worker}AMB-DG Algorithm (Worker Node)}
\end{algorithm}

\begin{algorithm}
\begin{algorithmic}[1]
		\ForAll {$t = 1, 2, \cdots$}
			\State initialize $b(t)=0, g(t)=0$.
			\ForAll {$i = 1,2 ,\cdots, n$}
				\State receive $b_i(t), g_i(t)$ from worker $i$
				\State $b(t)$ += $b_i(t)$
				\State $g(t)$ += $g_i(t)$
			\EndFor
			\State $z(t+1) = z(t) + \frac{1}{b(t)} g(t)$
			\State $w(t+1) = \arg \min_{w\in \mathcal{W}} \big\{ \langle w, z(t+1) \rangle + \frac{1}{\alpha(t+1)} \psi(w)\big\}$
			\State send $w(t+1)$ to all workers
		\EndFor
\end{algorithmic}

\caption{\label{alg:ambdg_master}AMB-DG Algorithm (Master Node)}
\end{algorithm}

\begin{IEEEbiography}[{\includegraphics[width=1in,height=1.25in,clip,keepaspectratio]{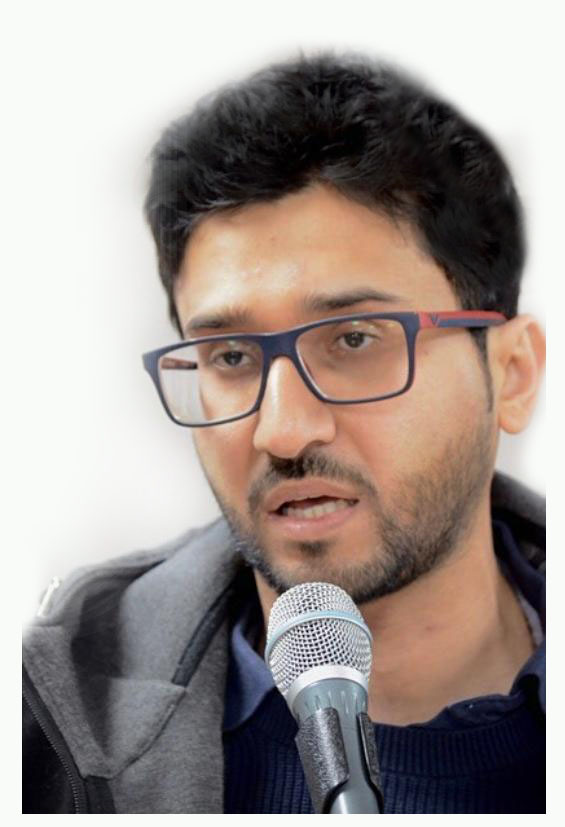}}]{Haider Al-Lawati} (Student Member, IEEE) received his B.Sc.E and M.Sc. degrees both in Engineering and Mathematics from Queen's University, Kingston, Canada in 2005 and 2007, respectively. He is currently a Ph.D. candidate at the Electrical and Computer Engnieering department at the University of Toronoto. Between 2008 and 2015, he worked in the telecom industry as an RF planning engineer, a project manager and a head of a department working on various mobile technologies such as GSM, WCDMA, LTE and WiMAX.
Throughout his academic and industrial careers, Mr. Al-Lawati received a number of scholarships and awards. His current research interests include distributed computing, distributed optimization, and machine learning.
\end{IEEEbiography}

\begin{IEEEbiography}[{\includegraphics[width=1in,height=1.25in,clip,keepaspectratio]{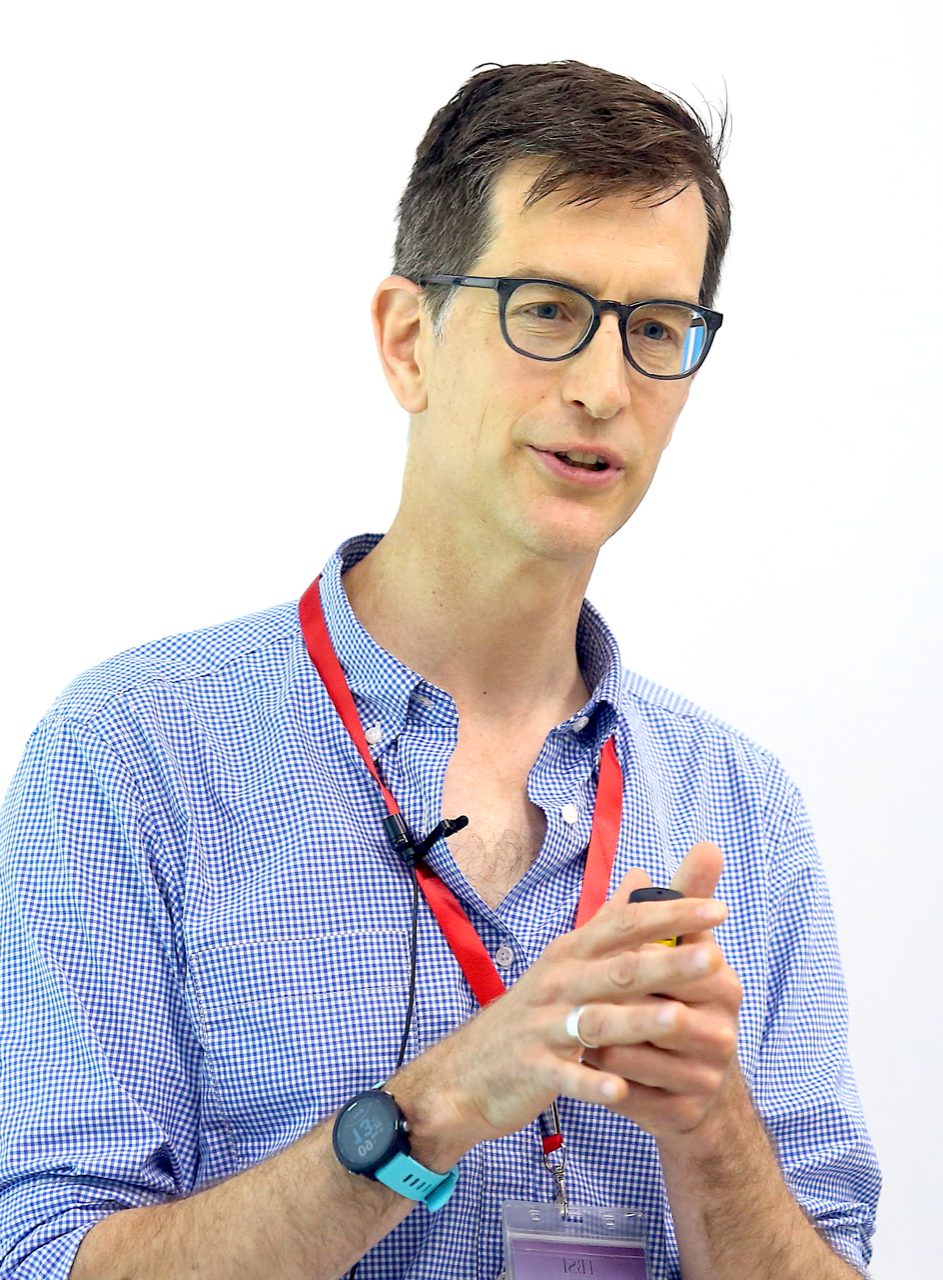}}]{Stark C. Draper} (Senior Member, IEEE) received the M.S. and Ph.D. degrees from the Massachusetts Institute of Technology (MIT), and the B.S. and B.A. degrees in electrical engineering and in history from Stanford University. He is a Professor of Electrical and Computer Engineering with the University of Toronto (UofT) and was an Associate Professor with the University of Wisconsin, Madison. As a Research Scientist he has worked with the Mitsubishi Electric Research Labs (MERL), Disney's Boston Research Lab, Arraycomm Inc., the C. S. Draper Laboratory, and Ktaadn Inc. He completed Postdocs with the UofT and at the University of California, Berkeley. His research interests include information theory, optimization, error-correction coding, security, and the application of tools and perspectives from these fields in communications, computing, learning, and astronomy.
Prof. Draper is a recipient of the NSERC Discovery Award, the NSF CAREER Award, the 2010 MERL President's Award, and teaching awards from the UofT, the University of Wisconsin, and MIT. He received an Intel Graduate Fellowship, Stanford's Frederick E. Terman Engineering Scholastic Award, and a U.S. State Department Fulbright Fellowship. He spent the 2019--2020 academic year on sabbatical at the Chinese University of Hong Kong, Shenzhen, and visiting the Canada-France-Hawaii Telescope (CFHT) in Hawai'i, USA.  He chairs the Machine Intelligence major at UofT, is a member of the IEEE Information Theory Society Board of Governors, and serves as the Faculty of Applied Science and Engineering representative on the UofT Governing Council. \end{IEEEbiography} 

\end{document}